\documentclass[preprint]{aastex7}

\usepackage{todonotes}
\usepackage{xspace}
\usepackage{subcaption}

\usepackage{graphicx}
\usepackage{amsmath}
\usepackage{listings}
\usepackage{tabularx}
\usepackage{float}
\lstdefinelanguage{json}{
    basicstyle=\normalfont\ttfamily,
    numbers=left,
    numberstyle=\scriptsize,
    stepnumber=1,
    numbersep=8pt,
    showstringspaces=false,
    breaklines=true,
    frame=lines
}

\definecolor{Gray}{gray}{0.7}
\definecolor{lightGray}{gray}{0.85}
\definecolor{darkblue}{rgb}{0.0,0.0,0.4}
\definecolor{darkgreen}{rgb}{0.0,0.4,0.0}
\definecolor{darkred}{rgb}{0.5,0.0,0.0}
\definecolor{grey}{rgb}{0.4,0.4,0.6}

\newcommand{\SynthPop}{{\sc SynthPop}}

\newcommand{\msun}{\,\mathrm{M_\odot}}

\newcommand{\pc}{\,\mathrm{pc}}
\newcommand{\matr}[1]{\mathbf{#1}}
\newcommand{\class}[1]{\textsc{#1}}

\newcommand{\srm}[1]{_{\rm {#1}}}

\shorttitle{SynthPop}
\shortauthors{Kl\"{u}ter \& Huston et al.}

\graphicspath{{./}{figures/}}

\begin{document}
 
\title{{\sc SynthPop}: A New Framework for Synthetic Milky Way Population Generation}

\author[0000-0002-3469-5133]{Jonas Kl\"{u}ter}
\altaffiliation{Co-first authors}
\affiliation{Department of Physics \& Astronomy, Louisiana State University, Baton Rouge, LA 70803, USA}
\email{jonas.klueter@web.de}

\author[0000-0003-4591-3201]{Macy J. Huston}
\altaffiliation{Co-first authors}
\affiliation{Astronomy Department, University of California, Berkeley, CA, 94720, USA}
\email[show]{mhuston@berkeley.edu}

\author{Abigail Aronica}
\affiliation{Department of Astronomy, The Ohio State University, Columbus, OH 43210, USA}
\email{abigail.aronica@gmail.com}

\author[0000-0001-9397-4768]{Samson A. Johnson}
\affiliation{Department of Astronomy, The Ohio State University, Columbus, OH 43210, USA}
\affiliation{NASA Jet Propulsion Laboratory, Pasadena, CA, 91109, USA}
\email{samson.a.johnson@gmail.com}

\author[0000-0001-7506-5640]{Matthew Penny}
\affiliation{Department of Physics \& Astronomy, Louisiana State University, Baton Rouge, LA 70803, USA}
\email{penny1@lsu.edu}

\author[0009-0002-1973-5229]{Marz Newman}
\affiliation{Department of Physics \& Astronomy, Louisiana State University, Baton Rouge, LA 70803, USA}
\email{mnewm25@lsu.edu}

\author[0000-0003-2872-9883]{Farzaneh Zohrabi}
\affiliation{Department of Physics \& Astronomy, Louisiana State University, Baton Rouge, LA 70803, USA}
\email{fzohra2@lsu.edu}

\author[0000-0003-4310-3440]{Alison L. Crisp}
\affiliation{Department of Astronomy, The Ohio State University, Columbus, OH 43210, USA}
\email{crisp.92@osu.edu}

\author[0000-0003-2558-1748]{Allison Chevis}
\affiliation{Department of Physics, Pennsylvania State University, State College, PA 16802, USA}
\email{allison@chevis.me}

\begin{abstract}

We present {\sc SynthPop}, a new open source, modular population synthesis Galactic modeling software to simulate catalogs of Milky Way stars along any sightline outward from the Sun. 
Motivated by a lack flexibility in existing Galactic models, {\sc SynthPop} is coded entirely in python, can be run standalone or as an imported module, and is configured by json files that allow different model components to be switched out as desired.
We describe the modular code structure, how the population generation process runs, and how to use the code. 
We also present model validation testing and known inaccuracies, and present an example of the code use, comparing Gaia data and the Gaia Universe Model Snapshot to a {\sc SynthPop} implementation. 
The code is available now via GitHub with ReadTheDocs documentation and can be installed via pip.

\end{abstract}


\section{Introduction}

Galactic population synthesis models have become an invaluable tool in various areas of astronomy involving the Milky Way. In their principal use case, they produce Monte Carlo forward model simulations of the complex mixture of stellar populations along a chosen line of sight, generating realizations not just of the number of stars, but also of the properties of the individual stars accounting for their evolutionary history. 
They bring together models of Galactic structure and kinematics, star formation history and chemical evolution, dust, and stellar evolution and atmospheres. In this paper we present an implementation of a new modular and flexible Galactic population synthesis code, {\sc SynthPop}\footnote{\url{https://github.com/synthpop-galaxy/synthpop}} \citep[v1.0.0;][]{synthpop_zenodo}, that addresses our perceived need for a code that can accurately reproduce observational data while also remaining flexible and open source, allowing users to adjust and improve upon it as new data and Galactic models become available.

Perhaps the earliest work on Galactic population synthesis models was conducted by Kate Brooks, who synthesized color magnitude diagrams from models of the vertical distribution of stars in the solar neighborhood; the work was presented by \citet{King1977}. More advanced models followed, notably, the widely-used Galactic model of \citet{Bahcall1980}, which synthesized the observable properties of stars from a combination of density functions and luminosity functions for the disk and halo. These early models, while able to create Monte Carlo realizations of stellar populations, were not true population synthesis models but are instead better thought of as samplers of empirically determined distributions.

The availability of public stellar evolution and atmosphere models enabled the development of true forward-model population synthesis models. Rather than drawing from an observed luminosity function, these models predicted the properties of stars in a population by first assigning them an age drawn from an age distribution or star formation rate (SFR) history, a mass from an initial mass function (IMF), and a metallicity from either an independent distribution or an age-metallicity relation (AMR). From these fundamental stellar properties, each star's current observational properties are determined by interpolation of stellar evolution models and atmospheres. The first to achieve this was the Hertzsprung-Russell Diagram Galactic Software Telescope model by \citet{Ng1995} that forward modeled the color-magnitude diagrams of a field in the Galactic bulge. 

The most widely used true population synthesis models today are the Besan{\c c}on \citep{Robin2003} and TRILEGAL models \citep{Girardi2005}, thanks to their key innovation of public web interfaces that enable anyone to generate realizations of synthetic stellar populations with ease. 

The Besan{\c c}on model was built over a series of papers introducing new Galactic components and model features.  Early versions of the model only considered stars from the disk and halo and, like the other early population synthesis models, relied on luminosity functions to determine the stellar populations' properties~\citep{Robin1986,Bienayme1987}. \citet{Haywood1997} updated the model to a true population synthesis model built from an SFR and IMF. After adding halo~\citep{Robin2000} and bulge~\citep{Picaud2004} populations, the Besan{\c c}on model was a complete population synthesis model of the Milky Way galaxy. It was made publicly available as a tool through a web interface~\citep{Robin2003}. In the two decades following, new versions of the Besan{\c c}on model have been released, updating various components to better match new data. Notably, the Gaia Universe Model Snapshot~\citep{Robin2012} model version was motivated by the then upcoming Gaia astrometry mission, and a version of it is available via the web or through a virtual observatory application programming interface\footnote{\url{https://model.obs-besancon.fr/}}. Despite the improvements in scriptable accessibility, the Besan{\c c}on model lacks the flexibility for the user to adjust the parameters of the model, likely to preserve its aim of dynamical self consistency.

The TRIdimensional modeL of thE GALaxy~\citep[TRILEGAL][]{Girardi2005} is a Galactic population synthesis model built upon the foundation of the Padova isochrones~\citep{Girardi2000}. As with the Besan{\c c}on model, its widespread adoption can be credited to its web interface\footnote{\url{http://stev.oapd.inaf.it/cgi-bin/trilegal}}. TRILEGAL's web interface is notably more flexible than Besan{\c c}on's, allowing the user to modify the parameters of all of the major components, addressing one of our goals of model flexibility, but its public interface lacks kinematic outputs and is not easily accessed via scripts.

At the outset of this project, the Galaxia code \citep[]{Sharma2011} came closest to meeting all of our requirements of a flexible population synthesis code that could be modified completely by the user. It is a public C++ code\footnote{\url{https://galaxia.sourceforge.net/}} that can use both analytical and $N$-body models to generate synthetic populations with extreme computational efficiency. However, it did not meet our flexibility requirement to enable easy implementation of a new model or modification of an existing model, as its analytic model components and their parameters are defined in C++ code that must be compiled. 

Other projects have built on these underlying codes to add flexibility in post-processing. Population Synthesis for Compact object Lensing Events \citep[PopSyCLE;][Abrams et al., in prep.]{Lam_2020, Rose2022} is a microlensing survey simulator, which begins with a Galaxia-generated catalog and adds compact objects and stellar multiplicity, generated via SPISEA \citep[Stellar Population Interface for Stellar Evolution and Atmospheres;][]{Hosek2020}. However, this does not allow for significant modification of the underlying populations.

We have built {\sc SynthPop} to address the need we perceived for a publicly available population synthesis code that can be easily modified, provides an interface that can be run via script, and ideally can allow comparisons between different models within a single framework. To achieve these goals, we built the package in python, which is likely the most widely used language in astronomy. This choice results in slower performance than a compiled language, but a minimal bar to code modification, and an easy way to import the package into other code. We have designed the package so that it can be modified either through the adjustment of parameter files or by adding new code modules to achieve results that are not enabled by existing tools.

This paper provides an overview of the {\sc SynthPop} code. We present the structure of the code in Section~\ref{sec:Modules} and the methods it adopts to generate synthetic populations in Section~\ref{sec:Generation}. Section~\ref{sec:How2Use} describes how to run and modify {\sc SynthPop} via json parameter files. Section~\ref{sec:known_issues} explains our process of validating individual components as well as known inaccuracies in the code. We provide a limited set of performance verification results for the full code and examples for use cases in Section~\ref{sec:examples}, but a full performance verification is beyond the scope of the paper, and performance for the desired task should be verified by the user for each model before relying on {\sc SynthPop}'s results, especially when modifying the models. Section~\ref{sec:summary} provides a discussion and summary of the paper. Two appendices describe the coordinate systems (Appendix~\ref{apx:coordinates}) and specialized isochrone interpolation scheme (Appendix~\ref{apx:charon}).

\section{Modules of the {\sc SynthPop} framework}
\label{sec:Modules}
A key goal of {\sc SynthPop} was to employ a modular design that would make modifying and swapping components simple, provided that new codes could conform their inputs and outputs to {\sc SynthPop}'s conventions. To achieve this and keep {\sc SynthPop} as flexible as possible, we split the key components of the star generation process into separate modules, each performing a single task. 

A subset of modules describe the model populations: \class{PopulationDensity}, \class{InitialMassFunction}, \class{Age}, \class{Metallicity} and \class{Kinematics}. Each population represents one of the structural stellar components of the Galaxy. These typically include a Galactic bulge and/or bar, thin disk (often specified as multiple age/metallicity-varied populations), thick disk, and halo. One may also include additional components like a nuclear stellar disk or nuclear star cluster.

Other modules control the estimation of stellar properties: the \class{Evolution} module, which consists of \class{EvolutionIsochrone} and \class{EvolutionInterpolator} sub-modules, and the \class{Extinction} module, which consists of \class{ExtinctionMap} and \class{ExtinctionLaw} sub-modules. 
Finally, \class{PostProcessing} modules can handle additional data treatment, e.g. applying specific cuts, calculating further stellar properties, or possibly to simulate survey observations. 

The instances for each of these modules can be specified independently. 
The only restrictions are the compatibility of the isochrone system and the isochrone interpolator and the extinction map wavelength being covered by the extinction law. 
From a software perspective, each module is implemented as an abstract parent class that defines the interface of the module, and the usable instances are subclasses of the parent class that implement the module's methods. In the following, we briefly describe the functionality of the modules. 

\subsection{Density profile}
\label{mod:Density}
The first property to describe a population is its density profile. 
It is used to estimate the mass or number of stars in a given volume. 
Typically, the density profile is given in respect to the evolved mass, but it may also use number or initial mass. 

Within the {\sc SynthPop} framework, the density calculation is handled by a \class{PopulationDensity} module. It estimates the density at a given cylindrical Galactocentric coordinate (\(r, \phi, \tilde{z})\), where \(\tilde{z}\) is height above the Galactic plane when accounting for the warp of the Milky Way. The full details of the convention we adopt for the Galactocentric coordinate systems, including that for the warp, are detailed in Appendix \ref{apx:coordinates}. 
The parent class also provides a numerical method to estimate the density gradient, which may be used later in calculating kinematics. 

In the initial release, we provide implemented subclasses for several density distribution from existing Galactic models: the Besan{\c c}on Model \citep{Robin2003}, its Gaia Universe Model Snapshot Variants \citep{Robin2012, Babusiaux2021}, and {\sc genstars} \citep{Koshimoto2021}. We also incorporate other Galactic structure studies, including the bulge density model of \citet{Cao2013} and nuclear stellar disk of \citet{Sormani2022}. 

\subsubsection{Optional Disk Flare}
In the outer regions of the Milky Way, the density profile of the disk might flare.
This can be accounted for in density profiles as a modified scale height \(h \cdot \kappa_{flare}\). 
To estimate the correction factor \(\kappa_{flare}\) we provide a function following \cite{Amores2017}:
\begin{equation}
    \kappa_{flare} = 
    \begin{cases}
		1  & \mbox{if } R \leq R_{flare} \\
		1 + \gamma \cdot (R-R_{flare}) & \mbox{if } R > R_{flare},
	\end{cases}
\end{equation}
and one following \cite{Mosenkov2021}:
\begin{equation}
    \kappa_{flare} = 
    \begin{cases}
		1  & \mbox{if } R \leq R_{flar} \\
		 e^{(R-R_{flare})/h_{flare}} & \mbox{if } R > R_{flare}.
	\end{cases}
\end{equation}
The corresponding function is selected by providing either \(\gamma_{flare}\) or \({h_{flare}}\).
Any \class{PopulationDensity} class where warp may be applied must include a call to the provided function for $\kappa_{flare}$ for flare to be implemented properly.

\subsection{Initial Mass Function}
\label{mod:IMF}
The next module is the Initial Mass function (IMF). This provides the first of three parameters ($m\srm{init}$) that characterize a star's properties and evolution in {\sc SynthPop}.
We provide modules for \citet{Kroupa2001}, \citet{Chabrier_2003}, and generic piece-wise power-law IMFs. 

To generate random initial stellar masses according to the IMF \(\xi(m)\), we use an inverse transformation method. This makes use of the cumulative distribution:
    \begin{equation}
        \label{eq:F_imf}
        F(m) = \int_{0}^m \xi(x) \,dx,
    \end{equation}
    and its inverse \(F^{-1}(p)\). 
First, values for \(F(m)\) are generated using a uniform distribution \(\mathcal{U}(F(m\srm{min}), F(m\srm{max}))\), where $m\srm{min}$ and $m\srm{max}$ are the minimum and maximum values assigned for initial stellar mass. 
The initial mass \(m_{\rm init}\) is then calculated via \(F^{-1}(F(m_{\rm init}))\). 
When implementing a new IMF, it is sufficient to specify  \(\xi(m)\). 
\(F(m)\) and \(F^{-1}(p)\) can be estimated numerically by {\sc SynthPop}. 
However, providing an analytical function for \(F(m)\) and \(F^{-1}(p)\) can improve run time.    

\subsection{Age and Metallicity}
\label{mod:Age_Met}
The {\sc SynthPop} framework uses isochrones so that stellar properties and magnitudes can be generated based on initial mass, age, and metallicity. Thus, the calculation of stellar properties is a 3-dimensional interpolation. {\sc SynthPop} generates age and metallicity values randomly from distributions. At present, {\sc SynthPop} uses only [Fe/H] to quantify metallicity. To approximate accounting for additional parameters, a user could, for example, implement a metallicity module with a correction formula to modify [Fe/H] based on $\alpha$-abundance \citep[e.g.][]{Yi2001, Salaris1993}.

For age, we provide modules for a single-age population, a uniform distribution across a selected range, an exponential distribution, and a Gaussian distribution. 
For metallicity, we provide a single-value module, a uniform distribution across a selected range, a Gaussian distribution, and a double Gaussian distribution.
The drawn stellar ages and positions are accessible within the metallicity drawing function, allowing a user to implement custom age-metallicity and metallicity-distance relations. For example, the provided \class{Gaussian\_Gradient} subclass for metallicity can vary metallicity with Galactocentric radius according to a custom input factor. A user may create module variants with other age- or position-dependent metallicity functions.

\subsection{Evolutionary Process}
\label{mod:Evolution}
One of the primary tasks of {\sc SynthPop} is to estimate stellar parameters and absolute magnitudes for stars given their initial parameters. For this purpose, {\sc SynthPop} interpolates across isochrone grids. 
This process is handled by the \class{Evolution} module, which contains an \class{EvolutionIsochrones} sub-module and an \class{EvolutionInterpolator} sub-module. 

The \class{EvolutionIsochrones} sub-module handles the loading and formatting process for a given isochrone system. This can include renaming columns or calculating additional columns based on the data given in the isochrones. 
Currently, we only provide a implementation of MIST \citep[MESA Isochrones and Stellar Tracks;][]{Dotter_2016, Choi_2016}, which were built from MESA \citep[Modules for Experiments in Stellar Astrophysics;][]{mesa1,mesa2,mesa3,mesa4,mesa5}. 
MIST provides isochrones on a regular metallicity-age grid for basic stellar properties and synthetic photometry for several magnitude systems. These cover initial masses from \(0.1-300 \msun\), [Fe/H] from -4.00 to +0.50, and $\log_{10}(\tau)$ from 5.0 to 10.3. 
The isochrones trace the evolution of some stars through part of the white dwarf cooling sequence and the rest to the post-Asymptotic Giant Branch (AGB) phase. 
Even though we only provide the MIST isochrone system, {\sc SynthPop} allows for the configuration of multiple isochrone systems for different mass ranges, allowing for coverage of further parameter space with additional modules. 
Future developments include adding isochrones for low-mass stars and brown dwarfs; currently, they may be generated if included in the IMF range but are not assigned evolved properties and photometry.
    
The \class{EvolutionInterpolator} sub-module interpolates among discrete points in the metallicity-age-initial mass grid to determine properties for each star.
As for the isochrones, it is possible to specify different interpolators for different mass ranges. 
This allows one to use a more precise but more computationally expensive interpolation close to the main sequence turnoff, or for the post-AGB phase. 
The default \class{CharonInterpolator} is described in Appendix \ref{apx:charon}. It uses linear interpolation in age and metallicity and cubic interpolation in initial masses, adjusted for the evolutionary phase of the star. We also provide a \class{LagrangeInterpolator} which excludes the evolutionary phase adjustments and thus performs worse in the stages where stellar evolution happens quickly (particularly the post-AGB phase). 

\subsection{Extinction}
\label{mod:Extinction}
In the conversion from absolute to apparent magnitudes, we use an extinction module, which estimates the extinction for a given location in the Galaxy in each given filter.
The \class{Extinction} module is  divided into two sub-modules: \class{ExtinctionMap} and \class{ExtinctionLaw}. A user may choose to exclude extinction by applying the \class{NoExtinction} subclass of \class{ExtinctionMap}.
    
    The \class{ExtinctionMap} module maps a value representing extinction throughout the Galaxy in 3D space. This value is the total extinction either at a reference spectral band \(A_{\lambda\srm{map}}\) or as a color excess E($\lambda\srm{map} - \lambda\srm{map2}$). 
    One implemented option is the use of the 2-dimensional or 3-dimensional extinction maps from the independent {\sc dustmaps} \citep{dustmaps} package \citep[currently available modules draw from:][]{Green2015,Green2018,Green2019,Burstein1982,Chen2014,Schlegel1998,Chiang2023, Delchambre2023, Sale2014, Edenhofer2024, Lenz2017,Marshall2006,Peek2010,PlanckCollaboration2014,PlanckCollaboration2016,Schlafly2011, Zucker2025}. Other implemented maps include the 2D map in the Galactic bulge direction from \citet{Surot2020} and the 3D map from \citet{Lallement2019} which extends 3 kpc in the Galactic plane and 0.4 kpc vertically. We suggest caution when using 2D maps, as they have no inherent distance dependence and dust is modeled as a single screen; a user may configure the distance at which the extinction screen is located. 
            
    The \class{ExtinctionLaw} module is then used to convert the extinction values (\(A_{\lambda_{\rm map}}\)) in the extinction map (given a reference wavelength \(\lambda_{\rm map}\)) to the extinction in some photometric filter (\(A_{\lambda}\)).
    Since the reference wavelength of the extinction law (\(\lambda_{\rm law}\)) is not necessarily the same as the reference wavelength in the extinction map (\(\lambda_{\rm map}\)), we estimate the 
    \(A_{\lambda}\) via: 
        \begin{equation}
    A_{\lambda} =  \frac{A_{\lambda}}{A_{\lambda_{\rm law}}} \cdot \frac{A_{\lambda_{\rm law}}}{A_{\lambda_{\rm map}}} \cdot A_{\lambda_{\rm map}}
    = \frac{f(\lambda)}{f(\lambda_{\rm map})} \cdot A_{\lambda_{\rm map}},
    \end{equation}
    where \(f(\lambda)\) is the extinction law. 
    If necessary, the module first converts color excess to a total extinction using, 
    \begin{equation}
        A_{\lambda\srm{map}} = \frac{1}{1-f(\lambda\srm{map2})/f(\lambda\srm{map})} \cdot E(\lambda\srm{map}-\lambda\srm{map2}).
    \end{equation}
    
    A user may specify different extinction laws for different wavelength ranges. We note that any extinction law used must cover the reference wavelength(s) for the extinction map. We provide implementations for the extinction laws specified by  
    \citet{Cardelli1989}, \citet{O'Donnell1994}, \citet{Fitzpatrick2009} \citet{Nishiyama2009}, \citet{Damineli2016}, \citet{Hosek2018} and \citet{Wang2019}. 

    Additionally, we developed the ``SODC'' (Surot-O’Donnell-Cardelli) extinction law to provide consistency between the \citet{O'Donnell1994} optical extinction law and the infrared reddening vector found by \citet{Surot2020}. We adopt the same form of extinction law as \citet{Cardelli1989},
    \begin{equation}
        \frac{A_\lambda}{A_V} = a(x) + \frac{b(x)}{R_V},
    \end{equation}
    where \(x = 1 / \lambda\) is the wavenumber, \(a(x)\) and \(b(x)\) are functions taking different forms in the visible (\(1.1 \, \mu\text{m}^{-1} \leq x \leq 3.3 \, \mu\text{m}^{-1}\)) and near-infrared (\(0.3 \, \mu\text{m}^{-1} \leq x \leq 1.1 \, \mu\text{m}^{-1}\)) regimes, and \(R_V = A_V / E(B - V)\) is the ratio of total to selective extinction. For optical wavelengths, we use the \citet{O'Donnell1994} coefficient values. For infrared wavelengths, we adopt the same form as the \citet{Cardelli1989} mean extinction law:
    \begin{equation}
    a(x) = a_1 x^\alpha, \quad b(x) = b_1 x^\alpha,
    \end{equation}
    and derived new values for \(a_1\), \(b_1\), and \(\alpha\) using the reddening vector found by \citet{Surot2020}, \(A_{K_s} = 0.422 \, E(J - K_s)\).
    The slope of the reddening vector determines the exponent \(\alpha = 2.255\), and we determined the coefficients \(a_1 = 0.539764\) and \(b_1 = -0.495567\) by requiring \(a(x)\) and \(b(x)\) to be continuous with the optical component across the boundary \(x = 1.1 \, \mu\text{m}^{-1}\), independent of \(R_V\).

\subsection{Kinematics}
\label{mod:Kinematics}
The \class{Kinematics} module is used to generate the Galactic space velocity ($u$, $v$, $w$), in the Galactocentric coordinate system (see Appendix \ref{apx:coordinates}). The generated velocities are also converted into galactic proper motions (\(\mu_{l^{*}},\, \mu_{b})\). 
Random velocities for stars are based on their location, and may also optionally depend on total stellar density or the stellar density gradient of the population(s). 
An example of a density-dependent kinematic model is the implementation of asymmetric drift in the Besan{\c c}on model \citep{Robin2003}. At present, \SynthPop{} does not provide information beyond stellar position and stellar mass density to the kinematic module. Additional factors like age- and metallicity- dependence are instead handled via population sets. For example, some Galactic models \citep[e.g.,][]{Robin2003,Koshimoto2021} split the Galactic disk into several components with different age and metallicity ranges, which also then have different velocity dispersion values.

\subsection{Post Processing}
\label{mod:PostProcessing}
Finally, we include a \class{PostProcessing} module for handling any additional data treatment after the stellar catalog has been generated.
This can be used for tasks like to re-filtering the catalog, adjusting data columns, and determining additional columns. We include a few simple modules including \class{GullsPostProcessing} to prepare catalogs for input into the {\sc gulls} microlensing simulator \citep{Penny2013,Penny2019}, \class{AdditionalCuts} to re-filter data on any existing column or pair of columns (e.g. for a color cut), and \class{ConvertMistMags} to convert MIST-generated magnitudes between systems (i.e., Vega, AB, ST).
It is possible to specify multiple \class{PostProcessing} instances to be run in sequence, as long as each is compatible with the output of the previously applied instances. 

For stars massive enough to burn carbon, the MIST isochrones only extend through the carbon burning phase of stellar evolution \citep{Choi_2016}. To more accurately account for the high-mass stars that have evolved beyond this point to become dark compact objects, we provide \class{ProcessDarkCompactObjects}, which removes the compact objects' photometry and assigns a more appropriate final mass. 
The scheme is based on the initial-final mass relations (IFMRs) implemented in PopSyCLE \citep[see][]{Rose2022}, with white dwarf masses based on \citet{Kalirai2008} and 3 different options for higher-mass remnants: Raithel18 \citep[based on][]{Raithel2018}, SukhboldN20 \citep[based on][]{Sukhbold2014,Sukhbold2016, Woosley2017, Woosley2020}, and Spera15 \citep[based on][]{Spera2015}.

\section{Generation Process}
\label{sec:Generation}
    In this section we describe the generation process for a {\sc SynthPop} catalog. Each stellar population (e.g. bulge, halo, disk, etc.) is generated separately, then merged before post-processing and output. 

\subsection{Star Counts}
    For each of the populations, {\sc SynthPop} generates a synthetic catalog of stars inside a cone extending outward from the Sun, defined by the Galactic latitude (\(l, \, b\)), the solid angle (\(\Omega\)), and maximum distance (\(R_{\rm max}\)). 
    First, {\sc SynthPop} splits the cone into slices along the distance axis with a thickness (\(\delta R\)) and some mean density (\(\overline{\rho}\)) within each slice volume ($V$). The mean density is calculated by sampling the density profile (see section \ref{mod:Density}) at \(N_{\rm mc}\) random locations per slice.
    In most of the cases, a small number of random locations is suitable to estimate the mean density within each slice, especially if the slice size (a function of $\delta R$ and $\Omega$) is small with respect to the scale length and height of the density profile.
    It then estimates the expected number of stars \(N_{\rm exp}\) to reside in the slice. 

    Each population's density profile is defined individually, and they need not be consistent in units (i.e. mass or number).
    If the density profile provides a number density, the expected number of stars is:
    \begin{equation}
        N_{exp}=\overline{\rho} \cdot V =\overline{\rho} \cdot 1/3 \cdot \Omega \cdot (D_{\rm outer}^{3} - D_{\rm inner}^3), 
    \end{equation}
    where \(D_{\rm inner\,/\,outer}\) are the inner and outer distances of the slice. 

    If the profile is defined as an initial (unevolved) stellar mass distribution, {\sc SynthPop} first estimates the average initial mass from the IMF (see Section \ref{mod:IMF}), $\overline{m}\srm{init}$. The expected number of stars is then:
    \begin{equation}
        N_{\rm exp}=\frac{\overline{\rho} \cdot V}{\overline{m}\srm{init}}.
    \end{equation}
    
    In the most common case, a population density is defined by the current masses of stars and remnant compact objects.
    In addition to $\overline{m}\srm{init}$, {\sc SynthPop} applies a mass loss correction factor 
    (\(a_{\rm mass\_loss} = \overline{m}_{\rm evol} / \overline{m}_{\rm init}\)), to transform between initial and evolved mass.
    We include a few options for how this factor is estimated: 
\begin{enumerate}
    \item evolving a test set of stars for each population and saving for all sightlines;
    \item evolving a test set of stars for each population individually for each sightlines;
    \item initially treating the population density as an initial mass density, evaluating the correction factor, and generating additional stars as needed (or following method 2 if that produces a larger set of test stars); or
    \item using an input, pre-calculated value for the mass loss factor or the average stellar evolved mass.
\end{enumerate}
    
    Option 4 is computationally fastest, and particularly useful for testing different density or kinematic profiles since the density does not affect the evolution. Option 1 is typically preferred for the balance in efficiency and accuracy; a user can define the test set size. Option 2 is a bit more accurate and less efficient, and option 3 is typically the most accurate but can get expensive for very large catalogs.
    The number of expected stars is then calculated via:
    \begin{equation}
        N_{exp} = \frac{\overline{\rho} \cdot V}{\overline{m}_{\rm init}\cdot a_{\rm mass\_loss}}.
    \end{equation}
    
    For any of the three conventions of population density, the number of stars generated in each slice (\(N_{\rm gen}\)) is then selected from a Poisson distribution with expectation value \(N_{\rm exp}\).

\subsection{Iterative Population Building Process}\label{sec:gen:iter}

    In some cases, it may be too computationally intense to produce and evolve all of these stars at once. For large populations, stars are generated in chunks of some configurable size. Thus, the following Subsections \ref{sec:gen:pos}-\ref{sec:gen:kine} may be executed iteratively several times. This also decreases memory usage for magnitude-limited catalogs by cutting stars dimmer than the limit during each iteration.

    The user may optionally select to skip low-mass stars for computational efficiency. In this case, stars which will not make the magnitude cut nor evolve beyond the main sequence can be accounted for in the total mass but excluded from the generation process.

    Velocities may be generated during the iterative population building process, or immediately following the completion of this process for all populations, depending on whether the full catalog information is needed for the kinematic module. For memory efficiency, the former option should be used unless the latter is required.

\subsection{Star Positions}\label{sec:gen:pos}
    Next, the position of each star is generated. We assume in this step that density is constant within a slice. Future work includes the removal of this assumption. {\sc SynthPop} first generates the offset to the line of sight (\(\theta\)) and the position angle of the offset (\(\Phi\)) following:
    \begin{align}
        \Theta &= \arccos(\mathcal{U}(\cos{\Theta_{max}},1))\\
        \Phi &= \mathcal{U}(0, 2\pi),
    \end{align}
    where \(\Theta_{max} = \arccos(1 - \Omega / (2\pi))\) is the opening angle of the cone. 
   The galactic longitudes and latitudes (\(l_{*},\,b_{*}\)) are then determined via:
    \begin{equation}
        \begin{pmatrix}
            \cos{b_{*}}\, \cos{l_{*}}  \\
            \cos{b_{*}}\,\sin{l_{*}}  \\
            \sin{b_{*}}
         \end{pmatrix} = \matr{R_{z}}(l)\matr{R_{y}}(b)
        \begin{pmatrix}
            \cos{(\Theta \cos{\Phi})}\cos{(\Theta \sin{\Phi})} \\
            \cos{(\Theta \cos{\Phi})}\sin{(\Theta \sin{\Phi})}\\
            \sin{(\Theta \cos {\Phi})}
        \end{pmatrix},
    \end{equation}
    where \(\matr{R_{y}}\) and \(\matr{R_{z}}\) are the rotation matrices around the y- and z- axes as defined in Equation \ref{eq:rot_matrix}.
    %
    The distances of the stars from the Sun \(d_{*}\) are generated using: 
    \begin{equation}
        d_{*} = \sqrt[3]{\mathcal{U}(D_{\rm inner}^{3}, D_{\rm outer}^{3})},
    \end{equation} 
    where $D_{\rm inner/outer}$ indicates the inner and outer distance boundaries of the current slice.
    
\subsection{Generating Stellar Parameters and Magnitudes}\label{sec:gen:mags}
    {\sc SynthPop} generates the initial mass, age and metallicity of each star according to their populations' distributions (see Sections \ref{mod:IMF} and \ref{mod:Age_Met}).

    The evolutionary process calculates the evolved mass, absolute magnitudes in a set of filters (\(M_{\lambda}\)), and a collection of additional stellar properties (e.g. temperature, stellar radii, surface gravity). These values are estimated via interpolation over the isochrone grid (see Section \ref{mod:Evolution}). 
    The set of magnitudes and stellar parameters can be defined in the configuration.
    {\sc SynthPop} can return either absolute magnitudes or reddened observed magnitudes. 
    Apparent magnitudes (\(m_{\lambda}\)) are calculated for each filter $\lambda$ via:
    \begin{equation}
        m_{\lambda} = M_{\lambda} + 5\cdot\log_{10} (d_{*}/10\mathrm{pc}) + A_{\lambda}(l_{*}, b_{*}, d_{*})
        \label{eq:mag_transform},
    \end{equation}
    where \(A_{x}(l_{*}, b_{*}, d_{*})\) is the total extinction for the given filter evaluated at the position each the stars as estimated by the \class{Extinction} module (see Section \ref{mod:Extinction}).
    
\subsection{Generating Velocities}\label{sec:gen:kine}

    The Galactocentric velocities of the stars are generated from the \class{Kinematics} module (see Section \ref{mod:Kinematics}). These are then transformed into proper motions and radial velocities via Equation \ref{eq:vel_trans}. 
    Radial velocities are provided both in the local standard of rest frame and the Solar system barycentric frame. We correct for the Solar reflex motion following \cite{Beaulieu_2000}:
    \begin{align}
            v_{\rm r,\,LSR}  = v_{\rm r} &+ v_{\rm LSR} \cdot \sin(l_{*})\cdot\sin(b_{*}) \\ \notag
        &+ v_{\rm pec} \cdot\left[ \sin(b_{*}) \sin(b_{\rm apex}) + \cos(b)\cdot\cos(b_{\rm apex})\cdot\cos(l_{*}-l_{\rm apex})\right],
    \end{align}
    where $v_{\rm LSR}$ is the circular velocity of the local standard of rest (LSR) at the Sun's position, $v_{\rm pec}$ is the Sun’s peculiar velocity relative to the LSR, and 
    $l_{\rm apex}$ and $b_{\rm apex}$ are the Galactic coordinates of the Solar apex.
A user may input custom Solar position and motion parameters in the model configuration\footnote{Because the observer's position can be manually set, \SynthPop~ could potentially be used to generate a catalog of stars from some position besides the Sun's, although the code and documentation will refer to the observer as the ``Sun.'' However, \SynthPop's {\sc Extinction} modules are calibrated for the Sun's position. A user may generate a catalog from a different Galactic perspective containing absolute magnitudes, with or without the distance modulus applied. This is not an intended use case for SynthPop, and we note that catalogs generated from a large Galactocentric distance may contain a very high number of stars, becoming very computationally expensive.}. Without these, we adopt the default values given in Table \ref{tab:solardefault}. 
    \begin{table}[h]
        \centering
        \begin{tabular}{|l|l|l|}
        \hline
            Parameter & Value & Source \\
            \hline
            ($x, y, z$)$_\odot$ & (-8.178, 0.0, 0.017) kpc & \citet{Karim2017, Abuter2019} \\
            ($u, v, w$)$_\odot$ & (12.9, 245.6, 7.78) km/s & \citet{Reid_2020} \\ 
            ($u, v, w$)$_{\rm LSR}$ & (1.8, 233.4, 0.53) km/s & \citet{Schonrich_2010} \\
            ($l, b$)$_{\rm apex}$ & (53$^\circ$, 25$^\circ$) & \citet{Beaulieu_2000,Mihalas1981}  \\
            \hline
        \end{tabular}
        \caption{Default Solar position and motion parameters.}
        \label{tab:solardefault}
    \end{table}

\subsection{Magnitude Limit \& Post-Processing}\label{sec:gen:post}
    After the iterative processes above are complete, the populations are compiled into a single catalog. Depending on several configuration parameters, the catalog may be cut by magnitude during the iterative population construction process to limit memory usage, or after all stars have been generated. The limiting magnitude and filter can be specified in the model configuration. 
    Lastly, the catalog is passed through any post-processing routines (see section \ref{mod:PostProcessing}).
    
\subsection{Output}
    \label{sub:Output}
    {\sc SynthPop} produces a synthetic catalog of stars with default and user-defined properties.
    The catalog contains several essential columns (listed in Table \ref{tab:output}), followed by optional properties from the isochrone interpolation (e.g. effective temperature, surface gravity) and magnitudes in the selected filters. 
    Optional properties and magnitudes can be specified within the model configuration file. The units of the additional columns and the magnitude systems are identical to the those in the isochrone system.
    If different units or magnitude systems not provided by the isochrones are needed, we recommend applying a post-processing script for the transformation (see Section \ref{mod:PostProcessing}). 
    
    {\sc SynthPop} stores the catalogs in a {\sc pandas} DataFrame, which can be saved in several formats including CSV, XML, HDF5 and Pickle. Alternatively, results can be saved in a fits format or as VOTable via the {\sc astropy.table} package. 
    The output format and location are be specified in the model configuration. 
    Scripts may also import {\sc SynthPop} as a python module, returning the DataFrame as the catalog generation function's output. 
    
    \begin{table}[h]
        \caption{This table lists the default output column names, the symbols used in this paper, the units for output values, and a brief description for all essential columns in the output table. Additionally, the resulting catalog may contain further user defined stellar parameters and magnitudes.}
        \centering
        \begin{tabular}{|c|c|c|c|l|}
        \hline
        \# &  Column name & symbol &  unit &  Description\\
        \hline
        1 & pop &  & & numerical integer indicating population \\
        2 & iMass & \(m_{init}\)& \(\msun\) & initial mass  \\
        3 & age & $\tau$ & \(\mathrm{Gyr}\) & age \\  
        4 & Fe/H\_initial & [Fe/H] & dex & initial metallicity \\  
        5 & Mass &  \(m_{evol}\) & \(\msun\) & evolved mass  \\
        6 & In\_Final\_Phase & & & beyond AGB-phase flag \\
        7 & Dist &  \(d\)  &\(\mathrm{kpc}\) & Sun-star distance  \\
        8 & l & \(l\) & degrees & Galactic longitude  \\
        9 & b & \(b\) & degrees & Galactic latitude  \\        
        10 & vr\_bc &\(v_{r}\)  & \(\mathrm{km/s}\) & radial velocity in Solar System barycentric frame   \\
        11 & mul & \(\mu_{l}\) & \(\mathrm{mas/yr}\) & proper motion in Galactic longitude  \\
        12 & mub & \(\mu_{b}\) & \(\mathrm{mas/yr}\) & proper motion in Galactic latitude  \\
        13 & x & \(x\) & \(\mathrm{kpc}\) & Galactocentric x-coordinate  \\
        14 & y & \(y\) & \(\mathrm{kpc}\) & Galactocentric y-coordinate  \\
        15 & z & \(z\) & \(\mathrm{kpc}\) & Galactocentric z-coordinate  \\
        16 & U & \(u\)& \(\mathrm{km/s}\) & x-component of the Galactocentric space velocity  \\
        17 & V & \(v\)& \(\mathrm{km/s}\) & y-component of the Galactocentric space velocity   \\
        18 & W & \(w\)& \(\mathrm{km/s}\) & z-component of the Galactocentric space velocity  \\
        19 & VR\_LSR & \(v_{r,\,lsr}\) & \(\mathrm{km/s}\) & radial velocity in the LSR frame  \\  
        20 & [A\_MAP]\footnote{[A\_MAP] is a placeholder for the extinction used in the extinction map, e.g. A\_Ks. It may instead be a color excess like E(B-V).} & \(A_{\lambda_{map}}\) &  & Extinction at the star's position \\ \hline
        \end{tabular}
        \label{tab:output}
    \end{table}

\section{Using {\sc SynthPop}}
\label{sec:How2Use}

{\sc SynthPop} is controlled by a hierarchy of json files. 
The configuration files (indicated by a `synthpop\_conf' file extension) control the overall settings of {\sc SynthPop}, and population files (indicated by a `popjson' file extension) define each individual \class{Population}. 

\subsection{Configuration Files}
\begin{figure}[ht!]
    \centering
    \includegraphics[width=12cm]{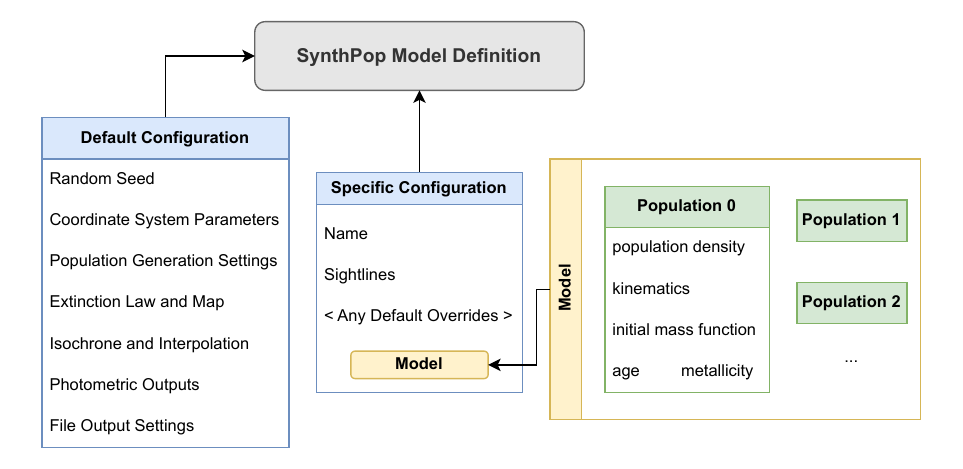}
    \caption{The structure of a {\sc SynthPop} model's definition. Two \textit{.synthpop\_conf} Configuration json files control the model settings. In these, a Model directory is selected, which contains some number of \textit{.popjson} json files specifying the Populations.}
    \label{fig:control_hierachy}
\end{figure}

The configuration files define several attributes which are global for the generation process. These include specifications for the lines of sight, the {\sc Extinction} and {\sc Evolution} modules, the selection of filters and stellar parameters, output formatting, and Solar position and velocity parameters. Figure \ref{fig:control_hierachy} shows the configuration and model structure.
Default settings are defined in a ``default'' configuration file. Any of these settings may be overwritten by a ``specific'' configuration file, which also must provide two mandatory parameters: the model name, and a name for the output. 
Working examples are provided for both. Specific configuration parameters may also be set as keyword arguments when running {\sc SynthPop} within a python script.

\subsection{Model \& Population File}
    For {\sc SynthPop}, a ``model'' is defined by a directory containing population files. Each population file describes a single population (e.g. the Galactic bulge or a disk component), and provides keyword settings for the selected IMF, density profile, age and metallicity distributions, and kinematics.
    Each population file is a json dictionary which provides the the name and/or filename of the selected module instances, as well as all arguments needed for their initialization. 
    Further, it may include the correction factor for the evolved mass or terms for the warp of the given populations. 
    In addition to working models, we provide a template of an empty population file for implementing new ones.

\section{Validation \& Performance}
\label{sec:validation}
We performed several tests to validate the generation process of SynthPop. 
The script for this validation is provided within the GitHub repository.

\subsection{Initial Properties} \label{sec:val:initial}
\begin{figure}[htb!]
        \centering
        \includegraphics[width=0.3\textwidth]{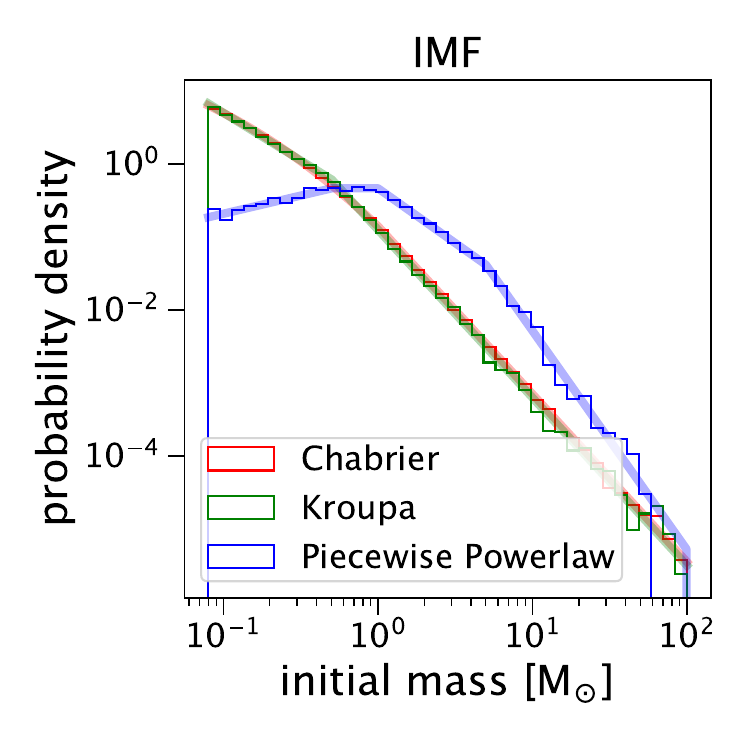}
        \includegraphics[width=0.3\textwidth]{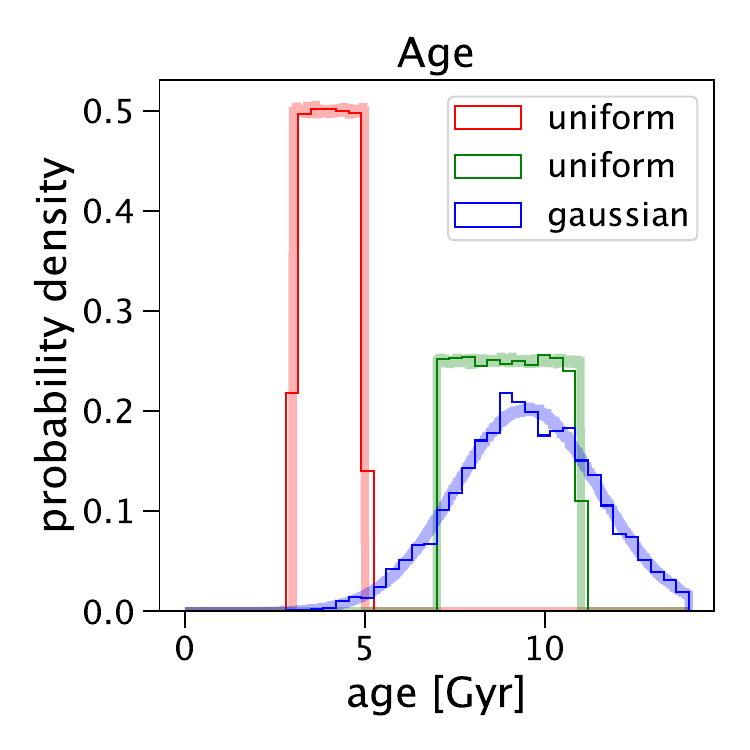}
        \includegraphics[width=0.3\textwidth]{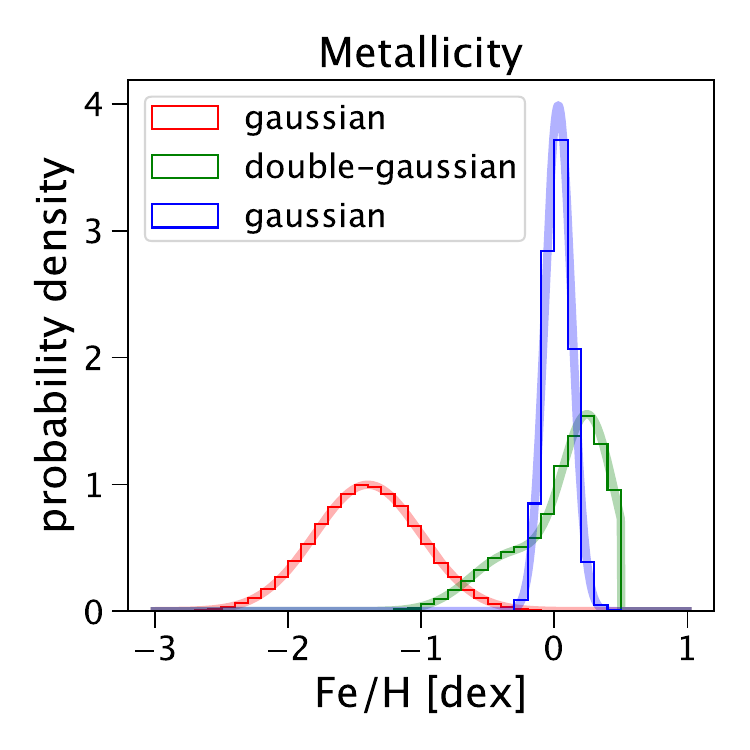}
        \caption{Comparison between input distributions (thick shadowed line) and distributions of generated stars (stepped line) for initial mass functions (left), age (center), and metallicity (right). 
        For IMFs, we show the Chabrier (red), Kroupa (green) and an arbitrary, unphysical piecewise power law (blue) distributions.
        For age, we show two different uniform distributions (red, green) and a Gaussian (blue).
        For metallicity, we show two single Gaussian distributions (red, blue), and a double Gaussian (green).
        In all cases, the generated stars follow the input distributions. }
        \label{fig:test_imf}
    \end{figure}

    We first investigate if the generated data follows the defined initial distributions. 
    To test this, we set up a validation model consisting of four populations, each with different IMFs, age distributions, metallicity distributions, and density profiles.
    We then generate the catalog along four lines of sight ($(l,b)$), two in key directions and two at random: (\(0^{\circ},\,0^{\circ})\), \((90^{\circ},\,0^{\circ})\), \((39.14^{\circ},\,8.53^{\circ})\), and \((12.17^{\circ},\,5.37^{\circ})\). 
    We selected the solid angles such that $\sim$100,000 stars are generated for each population along each line of sight. 
    Figure \ref{fig:test_imf} shows the resulting distributions of the initial stellar properties (stepped lines) for different distributions (thick lines). 
    In all cases, the distribution follows the likelihood function.
    Note that the parameters used for the piece-wise power law are not meant to represent a scientific IMF; it was defined to highlight that the generated stars follow the defined distribution. Figure \ref{fig:test_imf} also shows the equivalent and successful tests on age and metallicity distributions.

\subsection{Stellar Density \& Kinematics}

\begin{figure}[ht]
    \centering
    \includegraphics[width=0.23\textwidth]{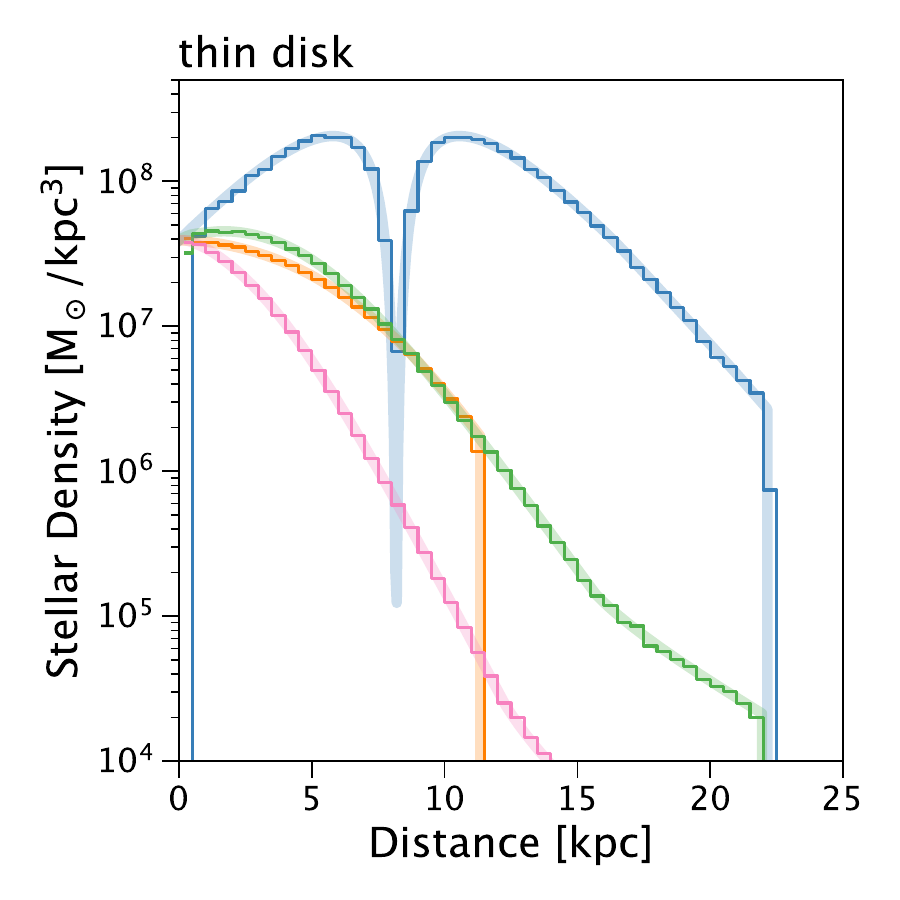}
    \includegraphics[width=0.23\textwidth]{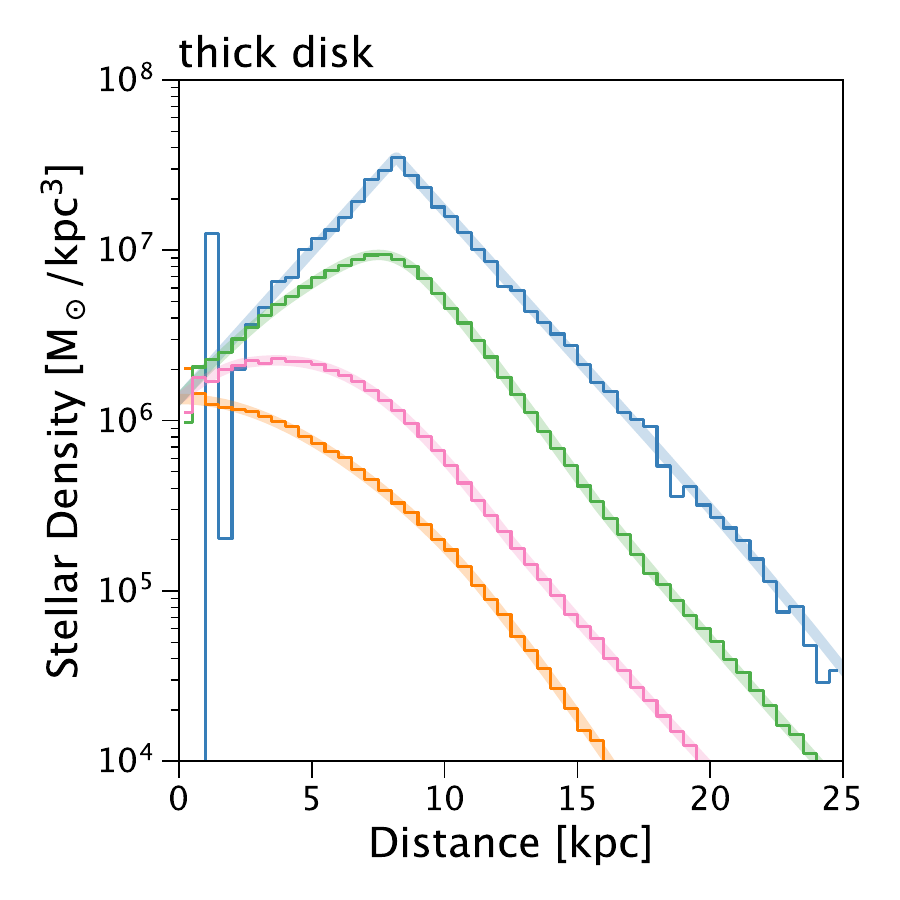}
    \includegraphics[width=0.23\textwidth]{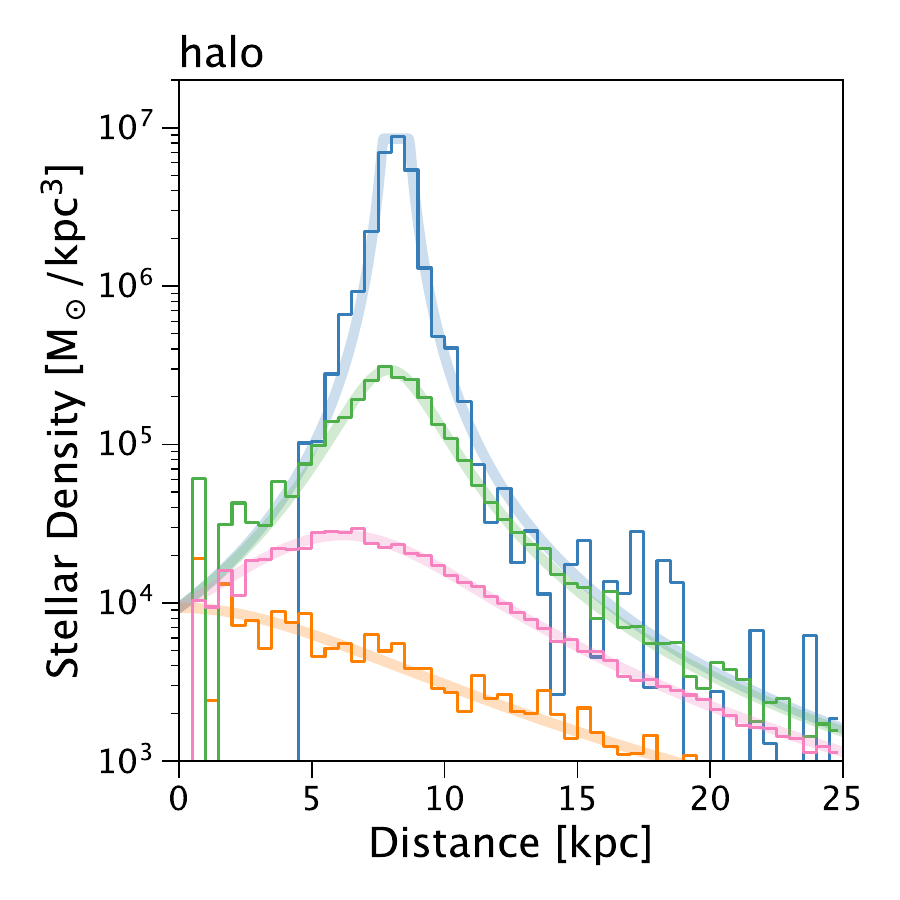}
    \includegraphics[width=0.23\textwidth]{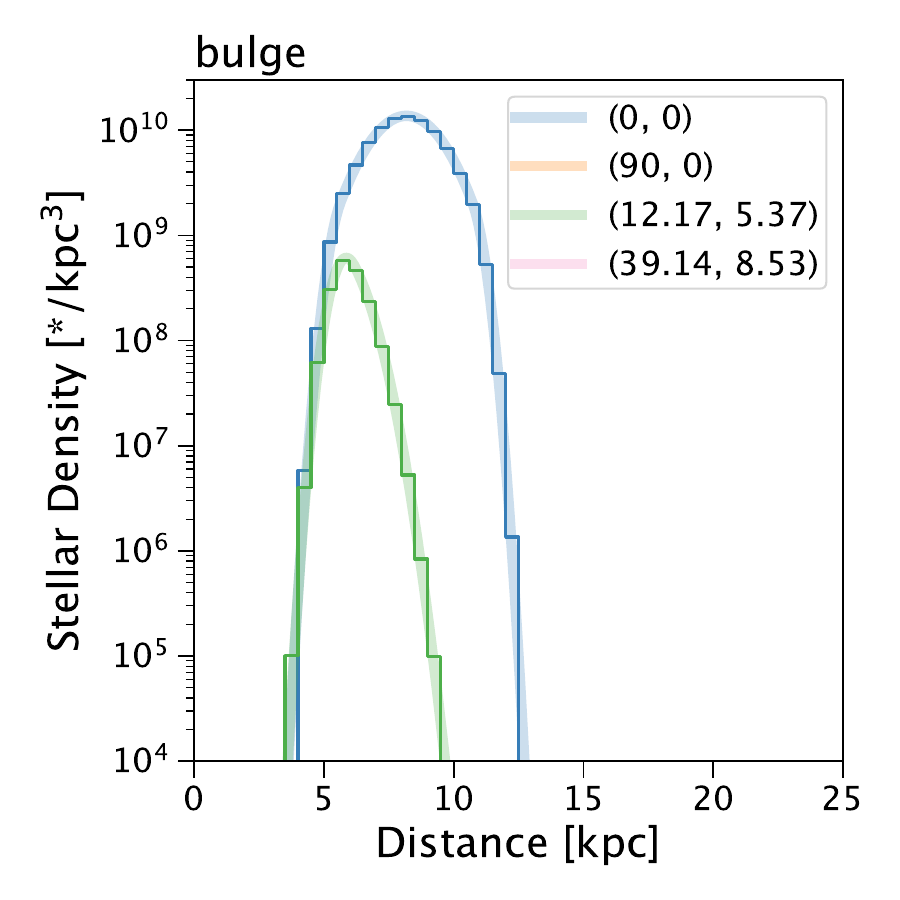}
     \caption{Stellar density for the (left to right) thin disk, thick disk, halo and bulge profiles used in the Besan{\c c}on \citep{Robin2003} Model.
     Each panel shows the stars produced (stepped line) and the specified distribution (shadowed line) along 4 different lines of sight. 
     Large scatter at small and large distances are due to small volumes and/or low densities resulting in very small numbers of stars generated. Note that the bulge population is defined by a number density rather than mass density. Two of the fields point too far from the Galactic center to contain any bulge stars.}
    \label{fig:test_density}
\end{figure}

To validate the star generation process in stellar density and kinematics, we again generate large catalogs along the same 4 lines of sight: (\(0^{\circ},\,0^{\circ})\), \((90^{\circ},\,0^{\circ})\), \((39.14^{\circ},\,8.53^{\circ})\), and \((12.17^{\circ},\,5.37^{\circ})\). This time, we use the \SynthPop~implementation of the Besan{\c c}on Model, as described by \citet{Robin2003}. Figure \ref{fig:test_density} shows the input density distribution for each of the populations, along with the effective density of generated stars in 0.5 kpc distance bins.
In all cases, the distribution follows the provided density profile. 

We used this same catalog set to validate the kinematics module, where the output velocities match the input distributions. For visual simplicity, we show a single population in the Galactic center direction $(l,b)=(0^\circ,0^\circ)$ in Figure \ref{fig:kinematic}. In this example, the Galactic cartesian velocities (see Appendix \ref{apx:coordinates}) follow visually clear distributions. The model follows a rotation curve for the circular velocity, and adjusts for population-dependent velocity dispersions and computed asymmetric drifts (see \citealp{Robin2003, Robin2004}).

\begin{figure}
    \centering
    \includegraphics[width=0.98\linewidth]{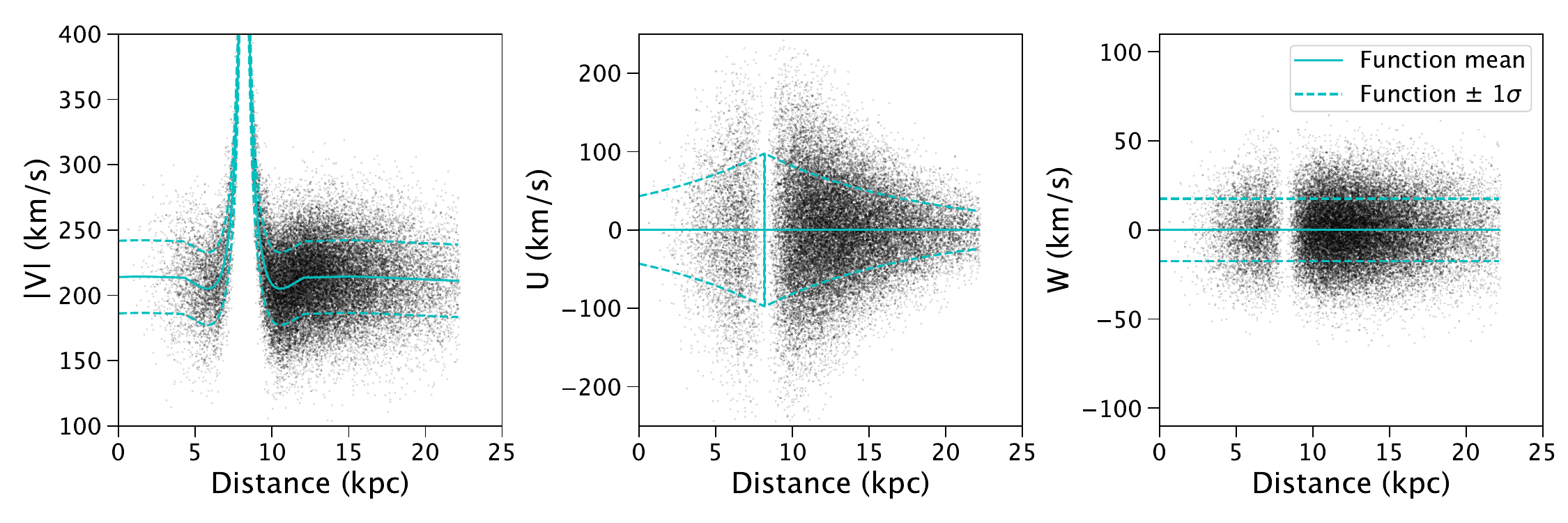}
    \caption{Galactic cartesian velocities of \SynthPop~simulated stars in the $(l,b)=(0^\circ,0^\circ)$ field belonging to the 7-10 Gyr thin disk sub-population. We limit the figure to a single population for clarity, as each population has its own velocity dispersions and asymmetric drift. The $|V|$ panel shows the absolute velocity in the direction of Galactic rotation, where true $V$ values are negative for the far disk. The $U$ panel shows the velocity component within the Galactic plane and perpendicular to the direction of Galactic rotation, where the velocity dispersion has a gradient. Finally, the $W$ panel shows velocity perpendicular to the Galactic plane. See Appendix \ref{apx:coordinates} for a full description of the coordinate system.}
    \label{fig:kinematic}
\end{figure}

\subsection{Stellar Evolution}

\begin{figure}
    \centering
    \includegraphics[height=7.5cm]{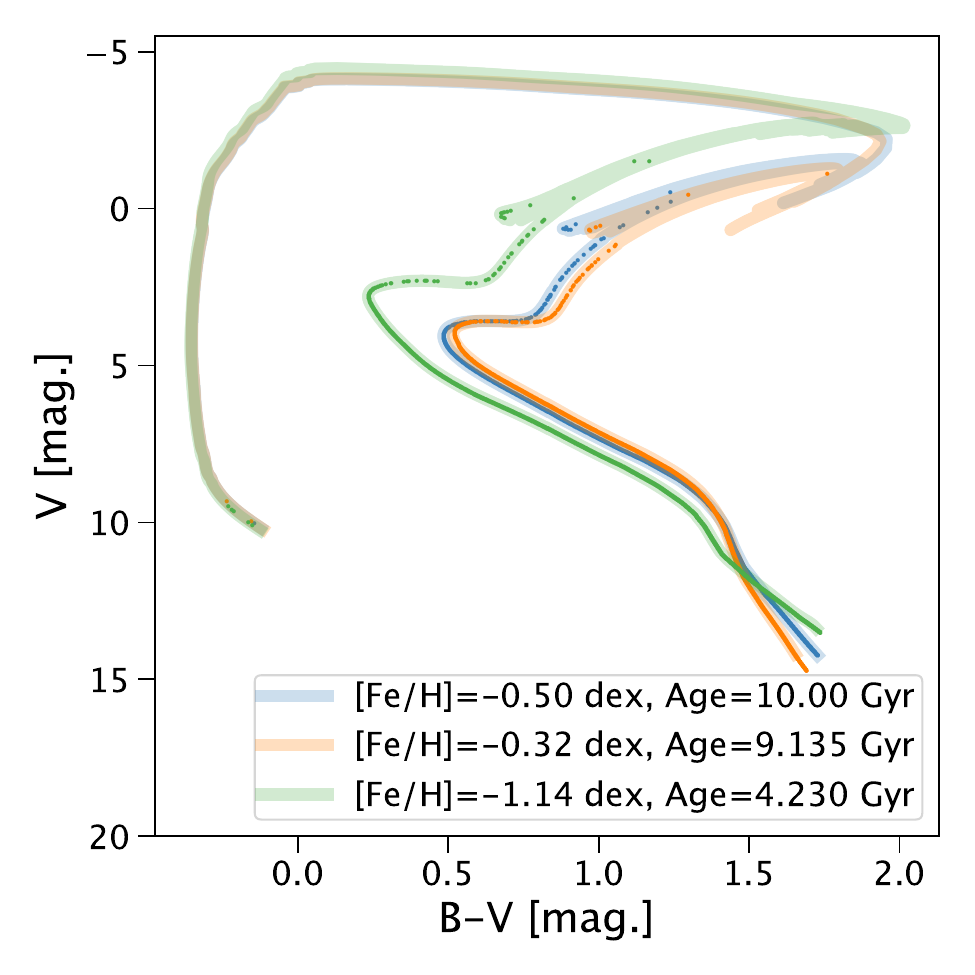}
    \caption{Comparison between the color-magnitude diagram generated by \SynthPop~(dots), and MIST web interpolator isochrones (thick shadow) for three different single-age, single-metallicity populations. B and V are shown in absolute Vega magnitudes in the Bessell filter system.}
    \label{fig:test_evolution}
\end{figure}

To validate the evolutionary process, we return to the non-physical validation model used in Section \ref{sec:val:initial}, examining three test populations each using a different single-age, single-metallicity model.  
The resulting absolute color-magnitude diagram is shown in Figure \ref{fig:test_evolution}. The three simulated populations are: [Fe/H] $=-0.5$ dex, age $= 10$ Gyr; $-0.32$ dex, $9.135$ Gyr; and $-1.14$ dex, $4.23$ Gyr. 
The first of these corresponds exactly to grid points in the MIST isochrone grid used for the interpolation. The other two populations require interpolation, and we use the MIST web interpolator for validation.
The magnitudes interpolated by SynthPop are in a good agreement with the MIST isochrones.
Note that the number of giant stars is boosted by using a non-physical IMF. 
For more details on the isochrone interpolation, see also Appendix \ref{apx:charon}.

.

\subsection{Performance}
\begin{figure}
    \centering
    \includegraphics[width=0.48\linewidth]{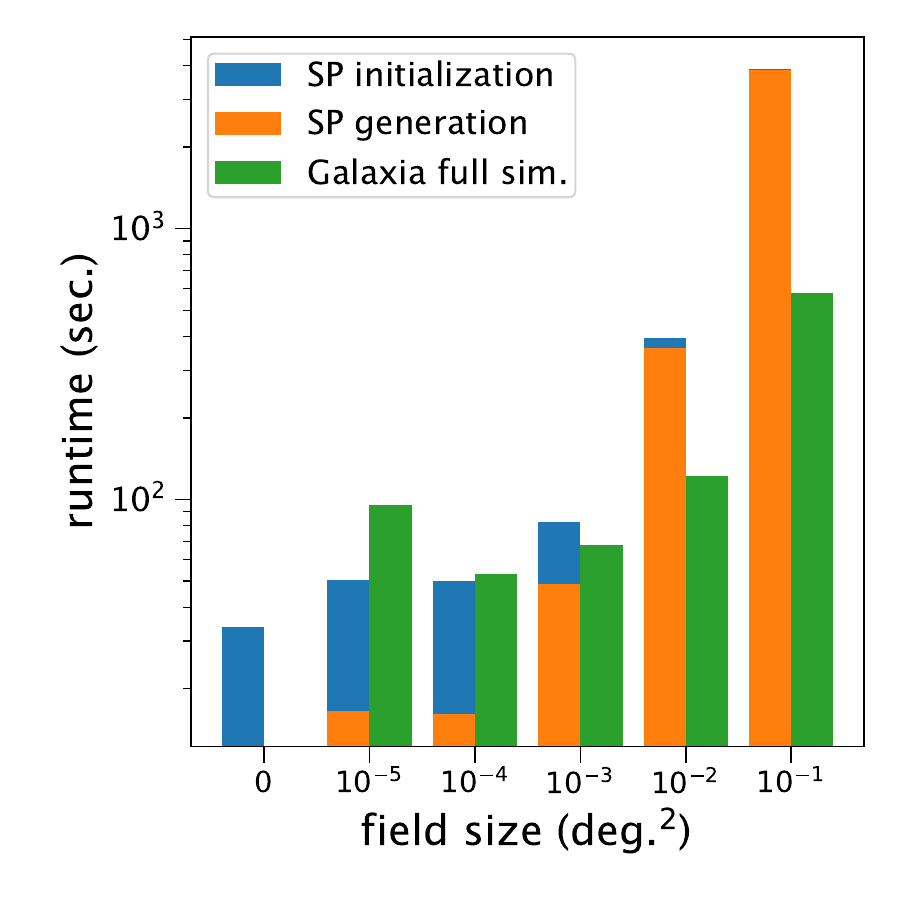}
    \includegraphics[width=0.48\linewidth]{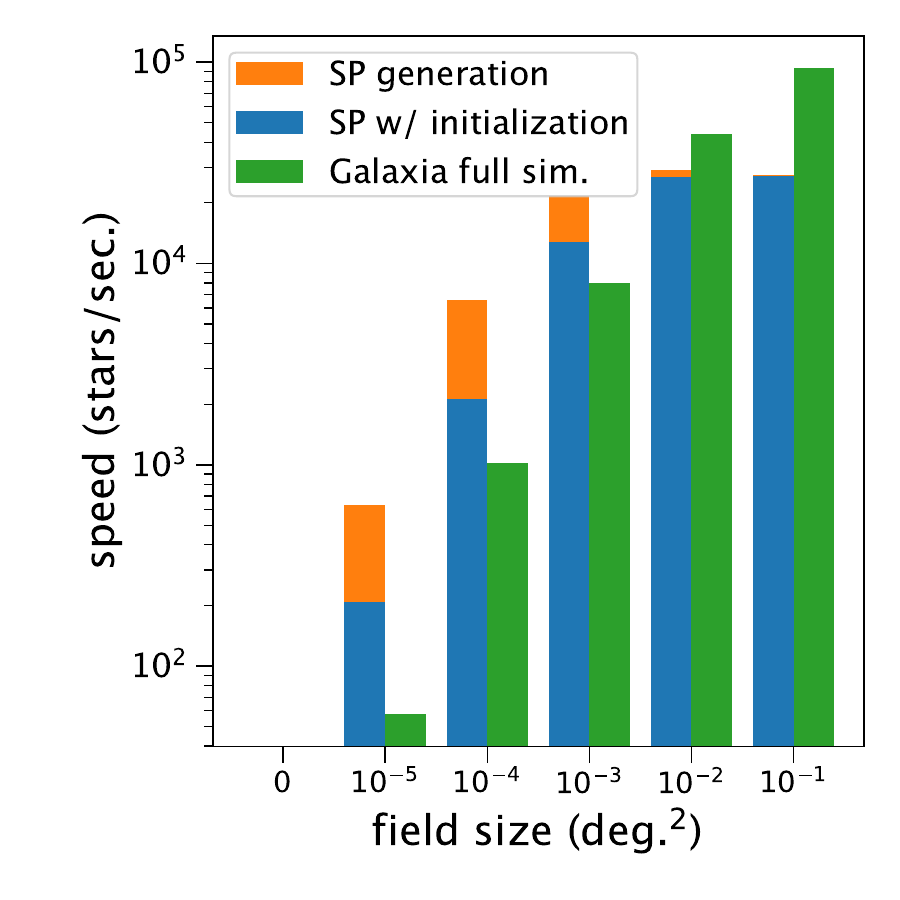}
    \caption{Runtime comparison of \SynthPop~and {\sc galaxia}. The left panel shows total runtime for a range of field sizes, and the right panel shows number of stars generated per second. In the left panel, the \SynthPop~values for the star generation process along are shown in orange, with initialization time shown in blue; in the right panel, these colors likewise indicate whether initialization was included in the total runtime. {\sc galaxia} has no separable initialization process and is shown in green.}
    \label{fig:runtime}
\end{figure}

We ran a performance comparison between {\sc galaxia} and \SynthPop, running both in serial on a 2017 iMac with a 4.2 GHz Quad-Core Intel Core i7 processor and a total of 64 GB 2400 MHz DDR4 memory. For {\sc galaxia}, we used the version available via GitHub\footnote{\url{https://github.com/jluastro/galaxia}}. Both codes were run with their Besan{\c c}on model implementations \citep{Robin2003}, directed arbitrarily toward $(l,b)=(-1.4^\circ,1.0^\circ)$. Figure \ref{fig:runtime} shows how the runtime of each varies with field size and number of stars. \SynthPop~ includes an initialization step, the runtime of which can vary, primarily depending on the extinction module set-up. Each individual {\sc galaxia} run includes file loading processes that \SynthPop~ does not need to do if already initialized. Thus, we show the \SynthPop~ runtimes with and without this initialization step, as it only needs to run once, regardless of whether one or many catalogs are then generated. For both, file saving time is included, with \SynthPop~using HDF5 (Hierarchical Data Format) and {\sc galaxia} using EBF (Efficient and Easy to use Binary Format).

We see that \SynthPop~performs faster for small catalogs ($\lesssim$1,000,000 stars). Isochrone interpolation is the most expensive part of the \SynthPop~generation process, and we see that computation time increases with star count for large catalogs. The {\sc galaxia} runtimes are much less strongly correlated to catalog size, and thus, run faster for large sizes ($\gtrsim1,000,000$ stars). Further optimization may improve the runtime of \SynthPop~in the future.

\section{Known Inaccuracies}
\label{sec:known_issues}

To keep the generation process flexible, {\sc SynthPop} includes a few assumption and shortcuts. These result in some minor inaccuracies.

\subsection{Constant density in a slice} 
    Currently we assume that the stellar number or mass density is constant within a slice. 
    At a distance of \(8\,\mathrm{kpc}\), a cone with a solid angle of \(1\,\mathrm{deg^{2}}\)
    has a diameter of \(150\,\mathrm{pc}\) which is on the same order as the default step size \(\delta R\) of 100 pc or typical resolution of extinction maps. For certain applications, instead of averaging over a large field, it may be optimal to combine multiple smaller fields. 

\subsection{Uncovered parameter space of the Isochrones}
    {\sc SynthPop} can generate stellar parameters for stars which are within the parameter space of the configured isochrone system. 
    To take their masses into account when estimating the number of stars, we need to estimate their evolved masses.
    For stars with masses below the minimum mass of the isochrone grid, we assume negligible mass loss (i.e. \(m_{evol} = m_{init}\)).
    For the MIST isochrones, these are stars with  \(m\srm{init} < 0.1 \msun\). 
    
    For stars above the maximum mass of an isochrone, we select the last evolved mass of the last isochrone grid point.
    For the MIST isochrones, this is typically a white dwarf with a mass of \(\sim0.7\msun\). This underestimates the masses for neutron stars or black holes. 
    The \class{ProcessDarkCompactObjects} \class{PostProcessing} class can improve the mass estimates for neutron stars and black holes, based a selection from 3 initial-final mass relations (adopting those used by \citet{Rose2022}). 
    However, this does not affect the number of stars generated. 
    
\subsection{Interpolation of Post-AGB Stars}
    At the end of their lifetime, stars go through several evolutionary phases on short timescales. 
    These are much shorter than the time steps in the default MIST Isochrones. 
    The \class{CharonInterpolator} (see Appendix \ref{apx:charon}) produces realistic properties and magnitudes for the large majority of initial masses. 
    However, large changes on very short timescales during the AGB and post-AGB phases are not reconstructed correctly, due to imperfect alignment of the evolutionary phases. 
    This is especially significant for metallicities [Fe/H] between $[-2.5, -2]$.
    While the MIST isochrones for [Fe/H]$=-2$ and ages above 2.5 Gyr end on the WD cooling sequence, the MIST isochrones for [Fe/H]$ =-2.5$ end on the red giant clump. This leads to an linear interpolation between two very different locations in the color magnitude diagram. 
    These stars are marked as stellar remnants outside of the interpolation grid. 

\section{Examples}
\label{sec:examples}

\subsection{Model comparison with Gaia DR3}
To show an example of {\sc SynthPop}'s functionality, we compare catalogs generated by {\sc SynthPop} with the Gaia Data Release 3 \citep[DR3,][]{GaiaCollaboration2023}, and the GAIA Universe Model Snapshot \citep{Babusiaux2021}.
We note that the main purpose of this section is to highlight the potential of {\sc SynthPop}, and it does not provide a new model of the Milky Way. Instead, we implemented a version of the Gaia Universe Model Snapshot (GUMS), as described in the Gaia DR3 Documentation\footnote{\href{https://gea.esac.esa.int/archive/documentation/GDR3/Data_processing/chap_cu2sim/sec_cu2UM/}{GAIA DR3 Documentation Section 2.2: Universe Model}}.

Because we are working from a different underlying framework and without all details about the GUMS model, there are some uncertainties and known inconsistencies between the models. The GUMS model provides a description of its parametrization and a queryable catalog\footnote{\href{https://gea.esac.esa.int/archive/documentation/GDR3/Gaia_archive/chap_datamodel/sec_dm_simulation_tables/ssec_dm_gaia_universe_model.html}{GAIA DR3 Documentation Section 20.10.2: gaia\_universe\_model}} of stars with Gaia magnitudes $G < 21$. Known bugs in the GUMS catalog affect the kinematics, which we exclude from this comparison.
Additionally, GUMS adds a 2-arm spiral structure to the thin disk that {\sc SynthPop} lacks. GUMS photometry uses the \citet{Sordo2011} spectral library, while {\sc SynthPop} uses the MIST isochrones \citep{mist1,mist2}. We assume the \citet{O'Donnell1994} extinction law with R$_V$=3.1, which may have a small effect on the conversion from the provided apparent magnitudes to absolute magnitudes.

To compare the output among the models and data, we select one arbitrary sight line near the Galactic center and one several tens of degrees away. Our ``inner field'' is at $(l, b) = (2, -2)^\circ$ with a cone/search radius of \(0.05^\circ\), and the ``outer field'' is $(l, b) = (30, 30)^\circ$ with a \(1^{\circ}\) radius. For both fields, we generate {\sc SynthPop} catalogs and query the GAIA DR3 database for the Gaia Source Catalog and the Gaia Universe Model.

    \begin{figure}
        \centering
        \begin{subfigure}[b]{0.48\textwidth}
         \includegraphics[width=\textwidth]{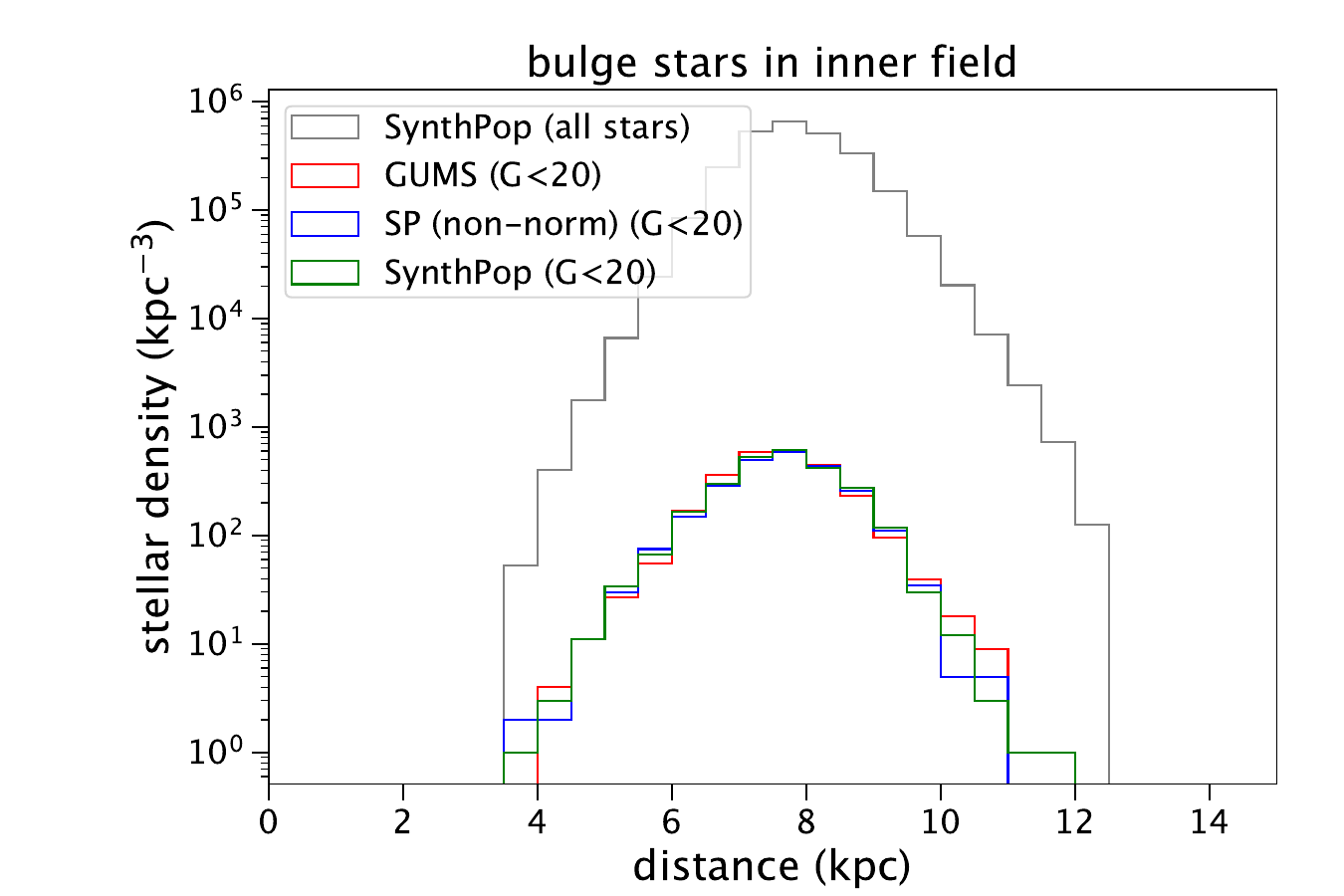}           
        \end{subfigure}
        \begin{subfigure}[b]{0.48\textwidth}
         \includegraphics[width=\textwidth]{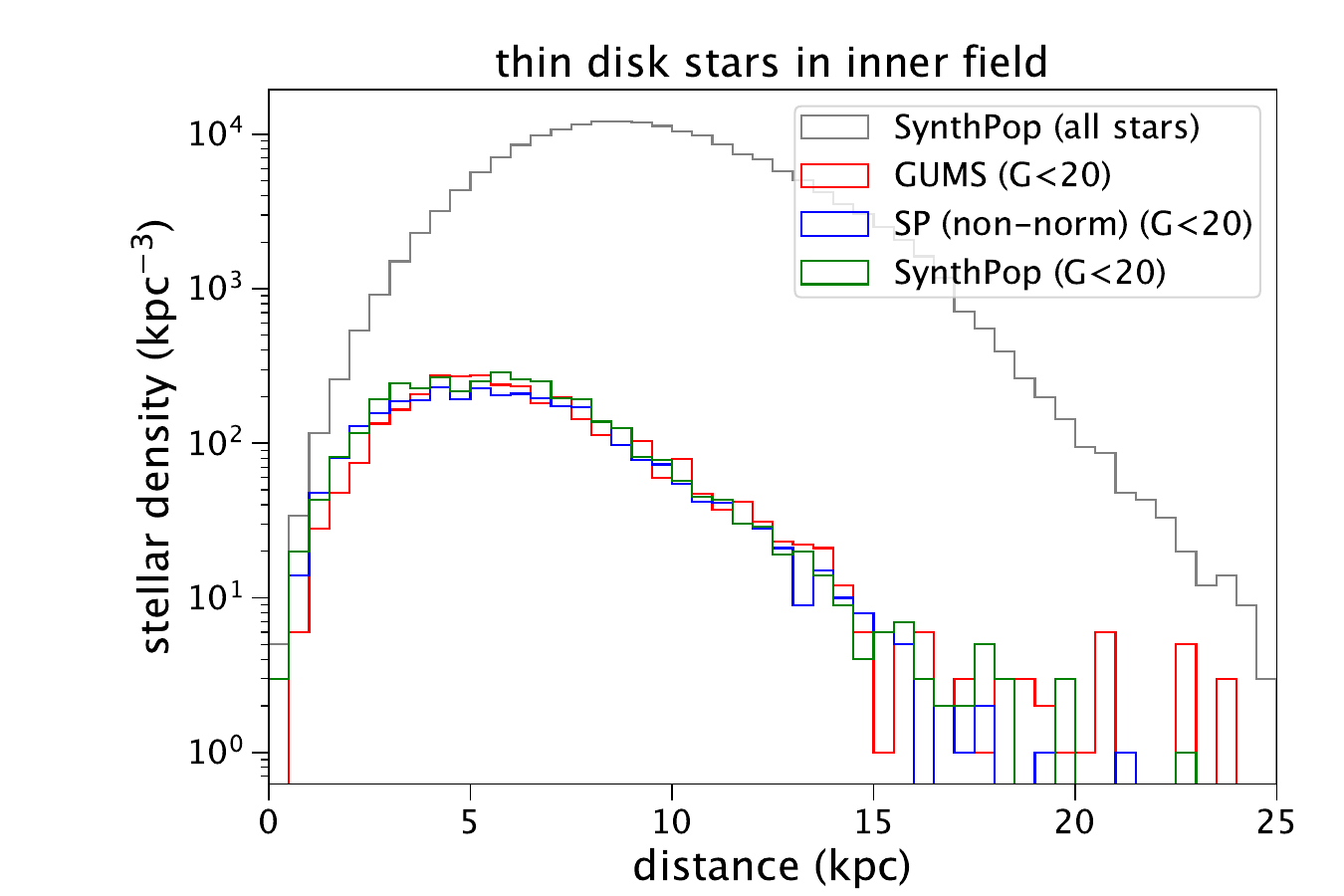}           
        \end{subfigure}
        \begin{subfigure}[b]{0.48\textwidth}
         \includegraphics[width=\textwidth]{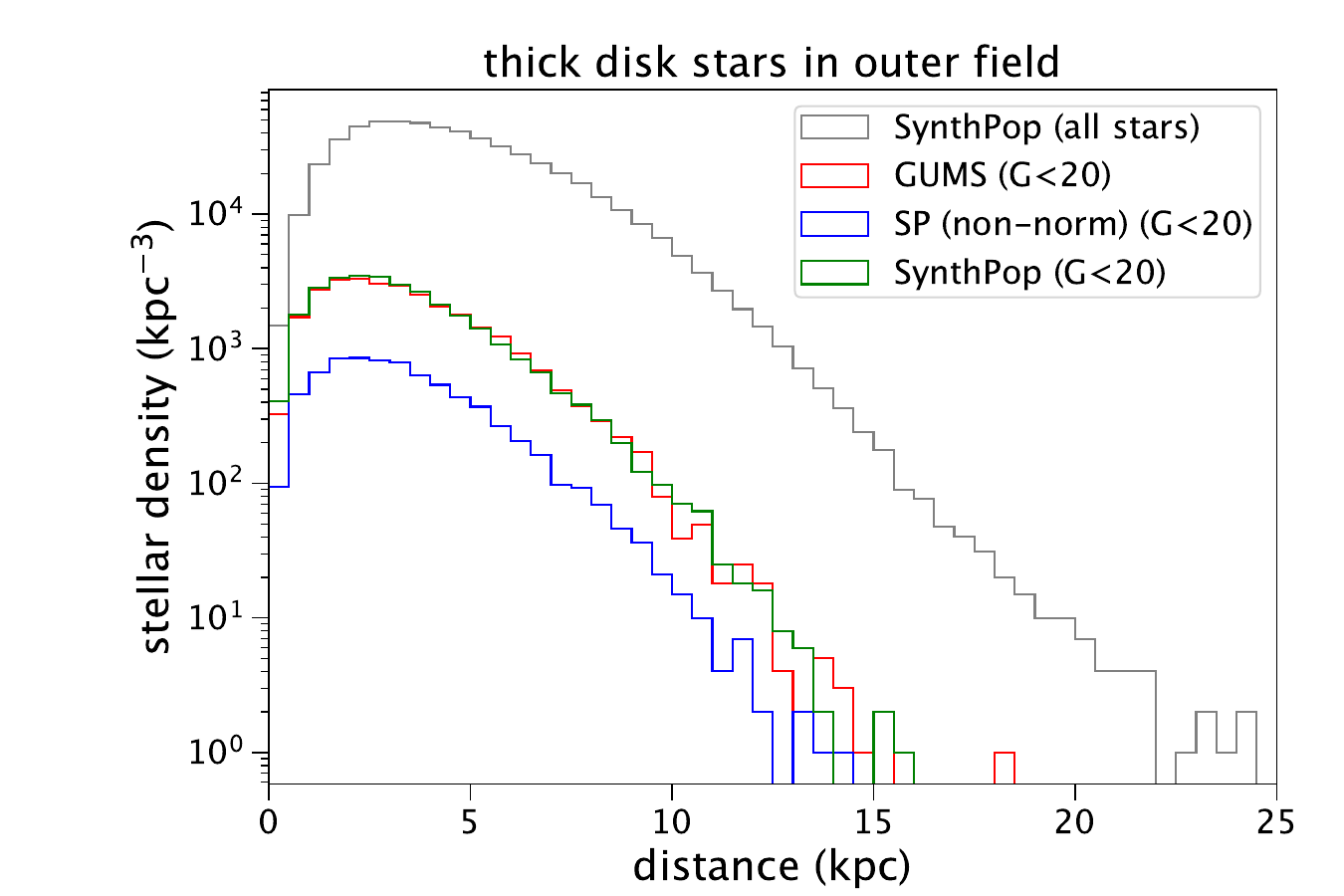}           
        \end{subfigure}
        \begin{subfigure}[b]{0.48\textwidth}
         \includegraphics[width=\textwidth]{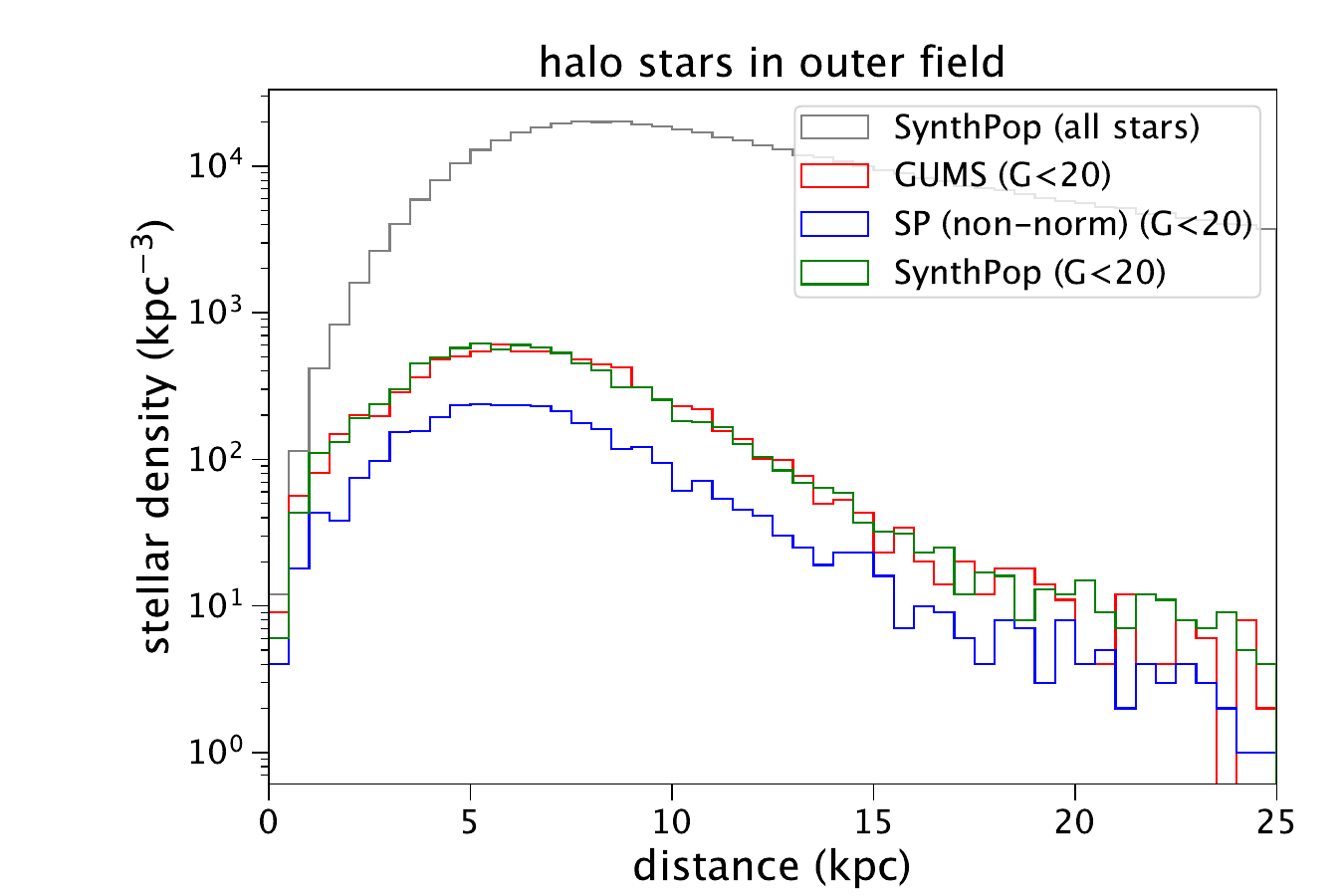}           
        \end{subfigure}
        \caption{Distance distribution for the bulge, thin disk, thick disk, and halo stellar populations. 
        The density of all stars generated by the {\sc SynthPop} model we adopt is shown in {\it grey}, reflecting the underlying total stellar density distribution for stars with $m>0.1$M$_\odot$. 
        The {\it green} and {\it blue} lines indicate the density for all {\sc SynthPop} stars brighter than 20mag in the Gaia G-band, with and without re-normalisation, respectively. The density estimated from the Gaia Universe Model catalog is shown in {\it red}.}
        \label{fig:dens_distribution}
    \end{figure}

Figure \ref{fig:dens_distribution} shows the density distributions for the bulge, thin disk, thick disk, and halo populations.
We compare the distributions of stars with \(G < 20\) generated via {\sc SynthPop} and GUMS, showing that the results do not match as well as expected for a re-creation of the model, illustrated by the difference between the non-renormalized {\sc SynthPop} stars and the GUMS stars. 
For the thick disk and halo, this difference disappears when we modify our model with re-normalized densities, assuming that the stellar mass density distribution provided by GUMS is for all stars except white dwarfs. 

For the thick disk and halo, this re-normalized {\sc SynthPop} model is in a good agreement with the data from the GUMS. We adopt this re-normalized version of the {\sc SynthPop} model as the {\sc SynthPop} model for the rest of this comparison.
Due to the younger age of the thin disk, the fraction of white dwarfs is much lower, and thus the renormalization does not have a significant effect. 
The bulge populations are parametrized by number rather than mass and are in good agreement without a renormalization.

\begin{figure}[H]
    \centering
    \includegraphics[width=\textwidth]{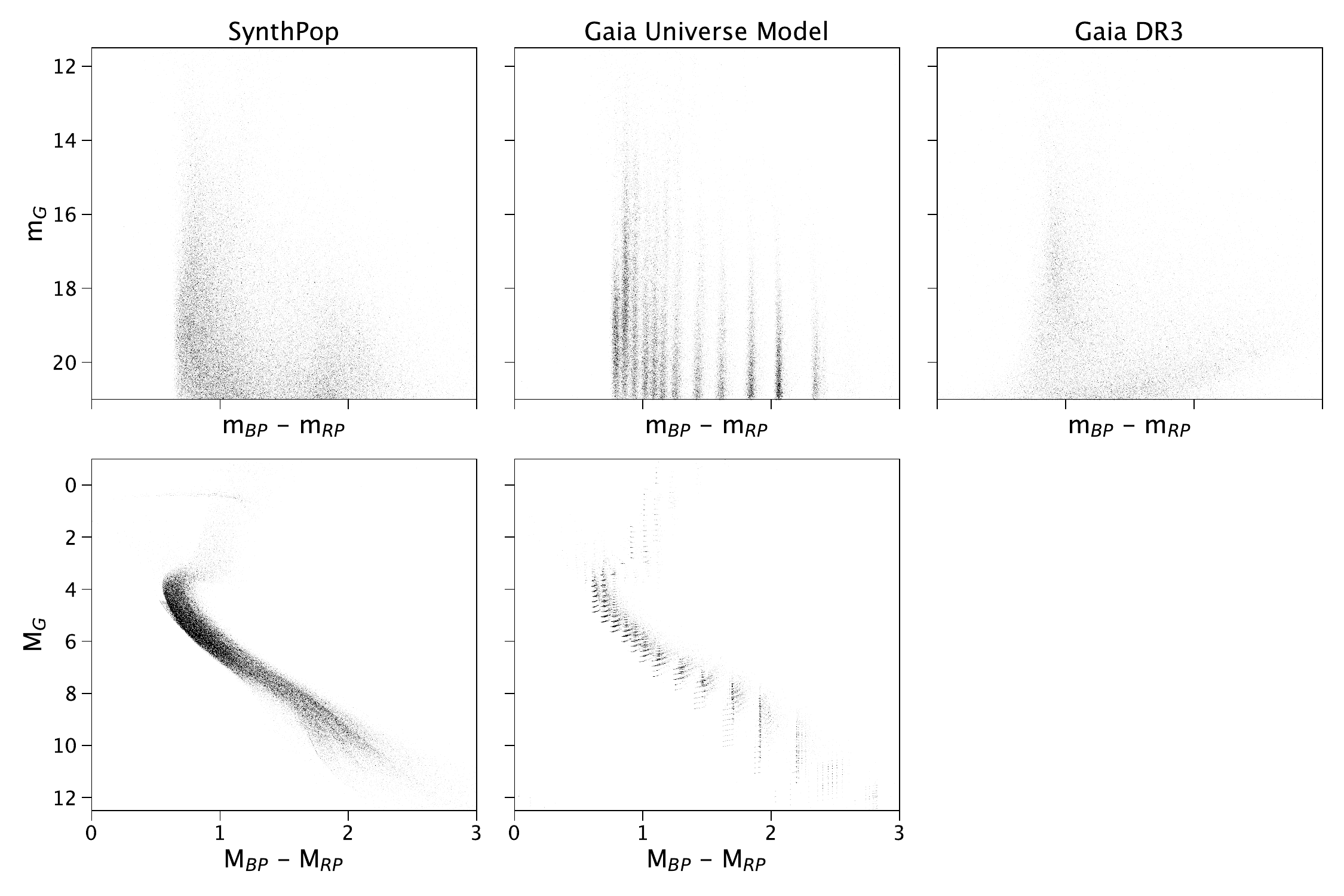} 
    \caption{Color-magnitude diagrams for {\sc SynthPop}, Gaia Universe Model and Gaia DR3. The top row is in observed magnitudes, and the bottom row is in absolute magnitudes.}
    \label{fig:comp_cmd}
\end{figure}

The resulting color-magnitude diagram (CMD) for the outer field is shown in Figure \ref{fig:comp_cmd}.
{\sc SynthPop} generates a smoother distribution for the absolute magnitudes, whereas absolute magnitudes from the GUMS form several clusters, which are then smeared out by extinction and distance to produce a more continuous apparent magnitude distribution. The resulting apparent distributions are very similar but still show striping in the GUMS colors.

An examination of the age and metallicity values provided by the GUMS catalogs shows that the individual stellar age values of the thin disk components are discretized to the middle values of each age bin. The SynthPop implementation spread the stellar ages over the age bins according to the exponentially decreasing star formation history provided in the GUMS documentation. This is likely to explain the difference in CMD smoothness.

\begin{figure}[H]
        \centering
        \begin{subfigure}[b]{0.48\textwidth}
        \includegraphics[width=\textwidth]{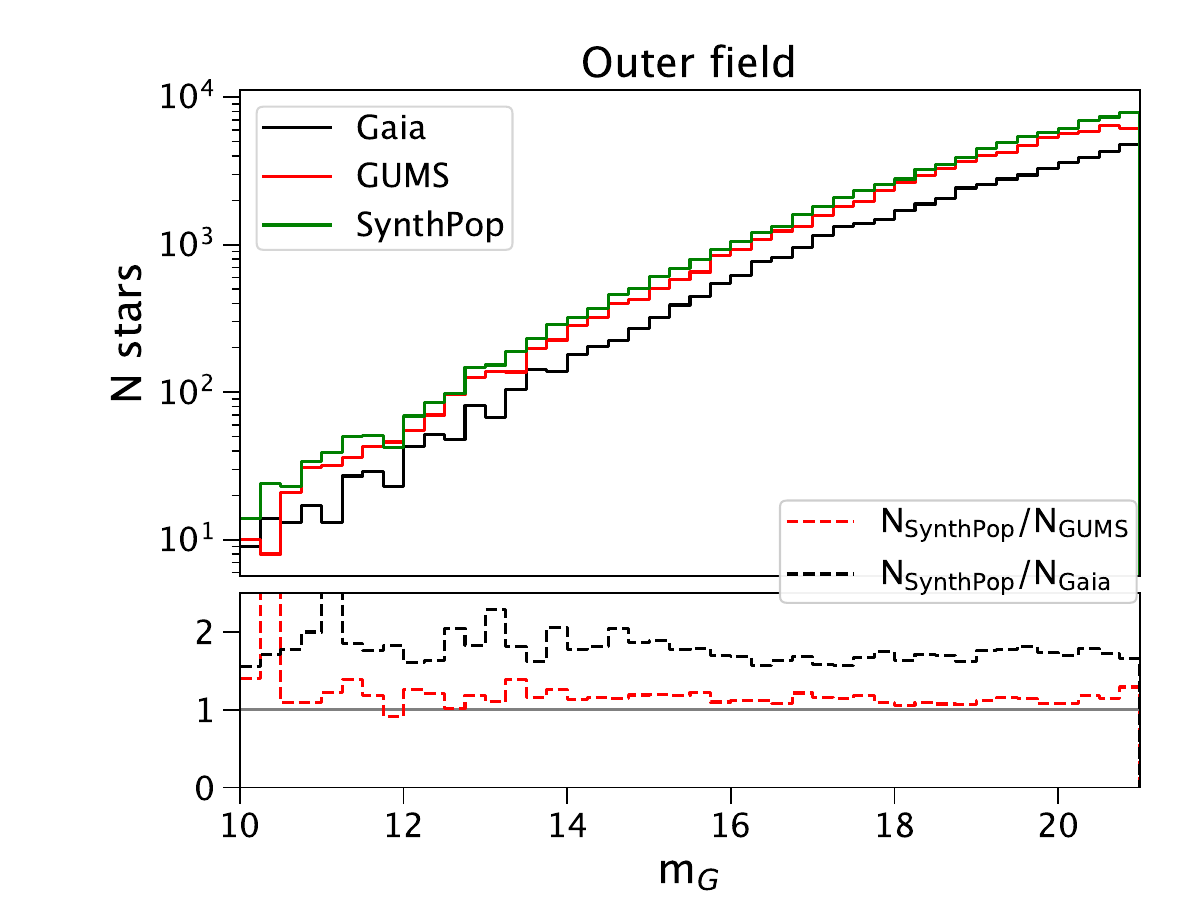}
        \end{subfigure}
        \begin{subfigure}[b]{0.48\textwidth}
        \includegraphics[width=\textwidth]{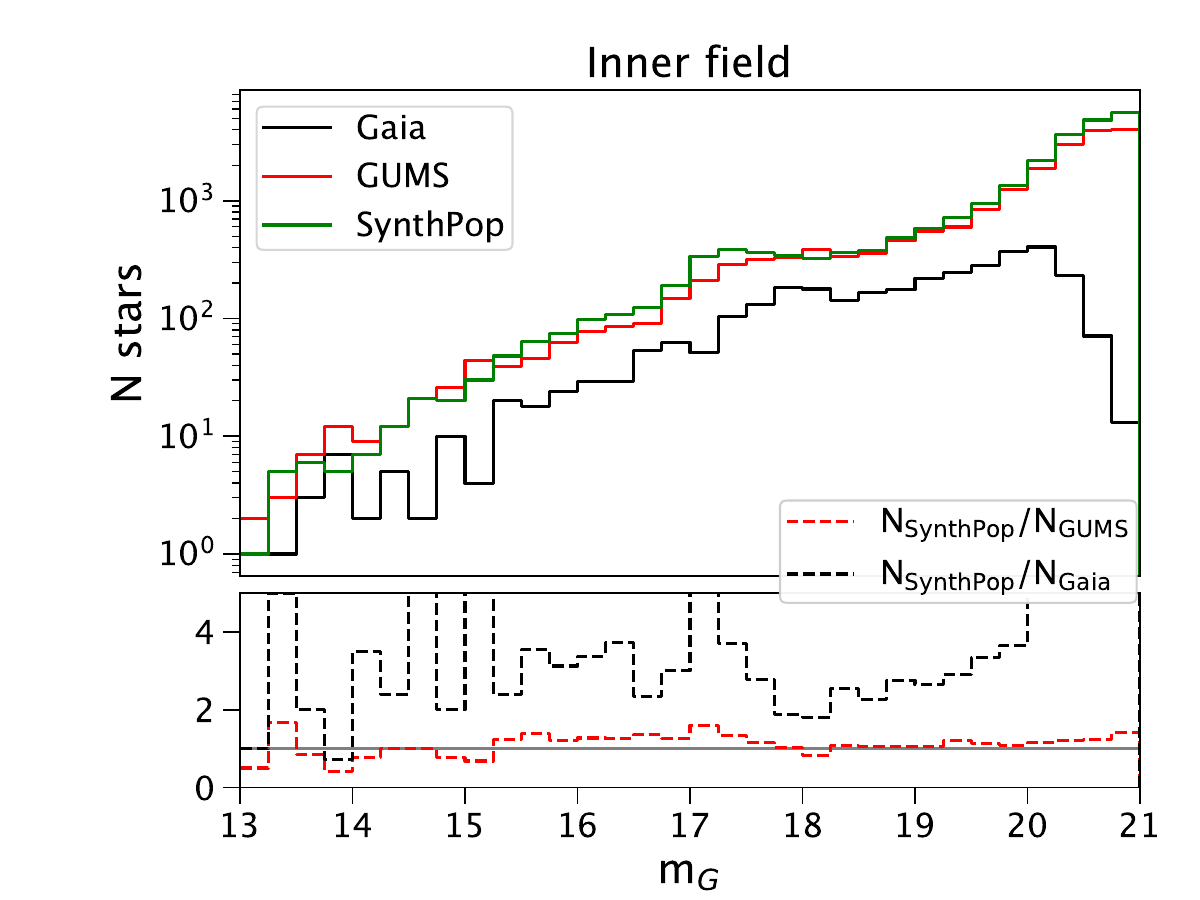}
        \end{subfigure}
        \caption{Luminosity function for the outer ({\it left}) and inner ({\it right}) Galactic fields. In {\it green}, the stars generated by {\sc SynthPop} are shown. 
        The {\it back} line shows the number of stars in Gaia DR3, and {\it red} shows the number of stars in the Gaia Universe Model catalog. The bottom panels give the ratio between the number of stars in {\sc SynthPop} and the Gaia DR3 ({\it black}) and the Gaia Universe Model ({\it red}).}
        \label{fig:gmag_hist}
    \end{figure}

Figure \ref{fig:gmag_hist} shows the G-band luminosity function for simulated stars from {\sc SynthPop}, simulated stars from GUMS, and real data from Gaia DR3. It shows both total counts for each data set and ratios of the {\sc SynthPop} simulated data set with respect to both the real Gaia stars and simulated GUMS stars. 

The {\sc SynthPop} output matches the GUMS stars well for both fields. Both models overpredict the true outer field Gaia population by $\sim$60\%. For the inner field, the mismatch between the models and Gaia stars is more egregious and varying by magnitude. This can at least in part be attributed to Gaia's poorer performance in crowded fields. For both fields, imperfect density profiles likely contribute to the model-data disagreement. This likely involves both issues with the large-scale populations as well as additional substructures not included. A full evaluation of these models against data is beyond the scope of this work.

Overall, we see that {\sc SynthPop} can produce stellar catalogs that approximately reproduce other models and/or real data. However, any {\sc SynthPop} implementation of another existing model will have some variation, based on underlying code structures and sometimes incomplete information.

\subsection{Microlensing Survey Preparation}
\citet{Kerins2009} first applied Galactic models to microlensing surveys, generating optical depth and event rate maps with the Besan{\c c}on Model \citep{Robin2003} and 3-dimensional extinction map from \citet{Marshall2006}. The output of microlensing surveys toward the Galactic center depends on the distributions of stars, dust, and stellar kinematics. This makes them an excellent test for models of the Galaxy. 

We developed {\sc SynthPop} with this in mind, creating a flexible, modular population synthesis Galactic modeling code. Huston et al. (in prep.) illustrate the development of a {\sc SynthPop} Model version that matches well to data, and apply the model to the Roman Galactic Bulge Time Domain Survey to explore the Galactic distributions of anticipated microlensing event lenses and sources. This model version is also being used as the Galactic model input for {\sc gulls} \citep{Penny2013, Penny2019} and PyLIMASS \citep{Bachelet2024} by the Roman Galactic Exoplanet Survey Project Infrastructure Team for updated exoplanet mass yield estimates  and field optimization (Terry et al., in prep.; Zohrabi et al., in prep.).

\section{Summary and Discussion}
\label{sec:summary}

In this paper we introduced {\sc SynthPop}, a modular framework to generate synthetic population models of the Milky Way. We outline the code base and detail its functionality, as well as demonstrate a sample validation against another Galactic Model. 
The code is accessible via GitHub\footnote{\url{https://github.com/synthpop-galaxy/synthpop}} with more detailed technical documentation available via ReadTheDocs\footnote{\url{https://synthpop.readthedocs.io/en/latest/}}.

{\sc SynthPop}'s modularity allows a user to set every component separately. Hence, it is much more flexible than other existing tools. This enables better comparison between models by only varying key components, allowing for more focused testing when developing new model distributions or combining elements from different models. 
Further, with the default implementation of the MIST Isochrone system, it is possible to include all magnitude systems and theoretical properties provided by MIST in pre-packaged isochrones.

The sacrifice required for this flexibility is computation time. 
Nevertheless, when testing {\sc SynthPop}, we found that it can generate about 50K stars per second using a single core, after initialization which takes several seconds, depending on the modules loaded. 
The interpolation is the most costly step of the catalog generation, expending approximately 60\% of the total computational time.
Therefore, the overhead can be reduced by implementing an interpolator that can determine whether a star is expected to meet the magnitude limit prior to the costly interpolation (see Section \ref{sec:gen:iter}).
Note that this computation time excludes saving the data in a human-readable format. 
Future development plans include further improving computational efficiency, potentially including multiprocessing.

To validate {\sc SynthPop}, we generated several simplified test populations.
In all cases, the resulting distribution in the catalog follows the defined distributions. The python script we used to generate the validation plots is included in the {\sc SynthPop} repository. 
However, the generation process of {\sc SynthPop} includes a few inaccuracies as described in Section \ref{sec:known_issues}. 
These are inaccuracies in the interpolation of stellar properties in short period between the AGB Phase and WD cooling sequence,
using the average density within a slice of the cone,
and effects due to limited isochrone parameter space coverage. 
None of these is expected to effect the results significantly, though this may depend on the intended use for the output catalogs. 
We aim to handle the known inaccuracies within future versions of {\sc SynthPop}.

Finally, we compared a {\sc SynthPop} model implementation of the Gaia Universe Model Snapshot (GUMS) with the original GUMS and Gaia DR3 data. 
When we renormalize our densities in {\sc SynthPop} for the thick disk and halo by assuming that white dwarfs are not included in the given densities, we find good agreement between our simulated catalogs and the Gaia Universe Model. 

We have not written {\sc SynthPop} with the intention of supplanting existing population synthesis models, which have been developed over years or decades. Instead, we intend it as a supplement or an enhancement to them for advanced use cases that require either large catalog simulations or the ability to tweak or alter parameters, models, or outputs. As such, we have attempted to implement versions of multiple existing population synthesis models in {\sc SynthPop}, and plan to add more with time. However, due to our own implementation choices and access to only what has been published about previous models, it is not possible to recreate each model with perfect accuracy. {\sc SynthPop}'s implementations of other models are therefore only an approximation. For this reason, users attempting to judge the performance of other models should ensure that {\sc SynthPop}'s implementation statistically reproduces the original models' output in each situation that they compare, and should rely on the original models' output for the final say. To appropriately recognize the contributions of the authors of the original models, any use of {\sc SynthPop} to generate catalogs of the implemented models should cite the relevant publications for those models in addition to citing this and any other {\sc SynthPop} software papers. 

The design, implementation, and testing of any software necessitates compromises, and as such, those we chose for {\sc SynthPop} have been shaped by our own research priorities. The primary driving use case for {\sc SynthPop} is to provide synthetic stellar population catalogs for microlensing simulations~\citep[e.g.,][]{Penny2013, Penny2019, Johnson2020}. As such, we have sought to optimize and test the code for large, deep catalogs in dense parts of the sky. Testing and validation has focused on the bulge and disk stellar populations viewed toward the Galactic bulge and plane. While we have tested the code over wider areas, those with use cases beyond the Galactic bulge area or relying on thick disk or halo stars should pay particular attention to the {\sc SynthPop} outputs and ensure they are reasonable. 

We welcome issues and pull requests to the GitHub repository\footnote{\url{https://github.com/synthpop-galaxy/synthpop}} to update and improve the code.
{\sc SynthPop} users are encouraged to share their configurations, scripts, and output with the user community via the {\sc SynthPop} Zenodo Community\footnote{\url{https://zenodo.org/communities/synthpop}}. 
This includes model configuration files, additional population files, new modules, scripts used to run {\sc SynthPop}, scripts to perform calculations on output catalogs, and the output catalogs themselves. 
Zenodo will aggregate these resources for sharing among the community and assign referenceable DOIs for each upload.         

\begin{acknowledgements}
    We appreciate conversations with N. Koshimoto in implementing the \citet{Koshimoto2021} model and \citet{Sormani2022} nuclear stellar disk into {\sc SynthPop}.
    We also thank the Roman Galactic Exoplanet Survey Project Infrastructure team members and others who helped improve the software through discussions, use during development, and feedback.

    J.K. acknowledges support from NASA award NNX16AC62G and the Gordon and Betty Moore Foundation award GBMF10467.
    M.J.H. acknowledges support from the Heising-Simons Foundation under grant No. 2022-3542. 
    Early work on this software by M.J.H. and A.A. was supported by the Ohio State University Summer Undergraduate Research Program.
    S.A.J.’s work was supported by NASA Grant 80NSSC24M0022 and an appointment to the NASA Postdoctoral Program at the NASA Jet Propulsion Laboratory, administered by Oak Ridge Associated Universities under contract with NASA. 
    M.T.P. acknowledges support from NASA awards NNX14AF63G, NNX16AC62G, 80NSSC24M0022, 80NSSC24K0881, and Louisiana Board of Regents Support Fund (RCS Award Contract Simple: LEQSF(2020-23)-RD-A-10). Early work by M.T.P. was performed in part under contract with the Jet Propulsion Laboratory (JPL) funded by NASA through the Sagan Fellowship Program executed by the NASA Exoplanet Science Institute.
    M.N. acknowledges support from NASA award 80NSSC24K0881.
    A.L.C. and F.Z. acknowledge support from NASA grant 80NSSC24M0022. 
    A.C.'s work was supported by the National Science Foundation under Grant Number 1852454.
    Portions of this research were conducted with high performance computational resources provided by Louisiana State University (\url{http://www.hpc.lsu.edu}).
    
\end{acknowledgements}

\software{NumPy \citep{numpy}, SciPy \citep{scipy}, pandas \citep{pandas1, pandas2}, Astropy \citep{astropy:2013, astropy:2018, astropy:2022}, dustmaps \citep{dustmaps}, Astroquery \citep{astroquery}}

\bibliography{references}{}
\bibliographystyle{aasjournalv7}

\appendix

\section{Coordinate Systems}
    \label{apx:coordinates}
    Typically, models of the Milky Way use Cartesian coordinates (\(x,\,y,\,z\)) or cylindrical coordinates (\(r,\,\phi,\,z\)), both with their origin at the center of the Galaxy, whereas observations are often described in spherical Galactic coordinates (\(l,\,b,\,d\)) centered on the location of the Sun. To connect observations and models, {\sc SynthPop} uses all three coordinate systems. 
    While the barycentric coordinate systems are standardized by the IAU, multiple conventions exist for coordinates systems centered on the origin of the Milky Way.
    This section describes our conventions of the Galactocentric coordinate systems, as well as our coordinate transformations.

    \begin{figure}[t!]
    \centering
        \includegraphics[width=\textwidth]{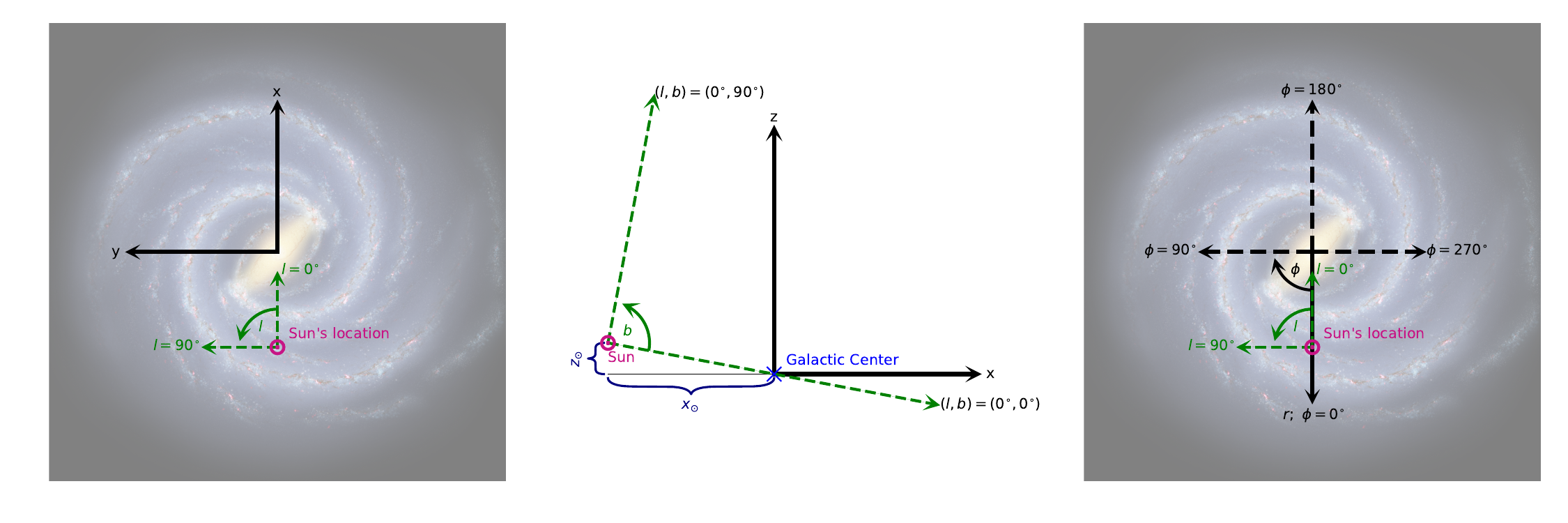}
        \caption{
        Comparison between the Galactocentric coordinates (black) and Galactic longitude and latitude systems (green). Left: Top view of the Milky Way with Cartesian coordinates. The \(z\)-axis points out of the page.
        Middle: Side view of the Milky Way with Cartesian coordinates. The $y$-axis points into the page ($x_0$, $z_0$ not to scale).
        Right: Top view of the Milky Way with cylindrical coordinates.
        The \(\tilde{z}\)-axis points out of the page.
        Underlying artist's concept of Milky Way from NASA / JPL-Caltech / R. Hurt (SSC-Caltech).
        }
        \label{fig:cartesian_coordinates}
    \end{figure} 

    \subsection{Cartesian Coordinates}
        We use Galactocentric Cartesian coordinates \((x,\,y,\,z)\) and the reference frame for the Galactic velocities  \((u, v, w)\). 
        The convention (see Figure \ref{fig:cartesian_coordinates}) is defined by its x-axis pointing along the Galactic plane, opposite the Sun's direction and its z-axis pointing toward the Galactic North pole.
        The y-axis completes the right-handed coordinate system and points roughly toward the Galactic coordinates \((l,b) = (90^{\circ}, 0^{\circ})\). At the position of the Sun, $y$ increases in the direction of Galactic rotation. 
        That means that the position of the Sun is \((x,\,y,\,z)_{\odot} \sim (-8000, 0, 10)\pc\), 
        where $z_\odot$ is the distance above the Galactic plane, 
        and \(x_{\odot}=\sqrt{d_{\rm GC}^2-z_{\odot}^2}\simeq d_{\rm GC}\), given the distance to the Galactic center $d_{\rm GC}$. Default values for the Sun's position are listed in Table \ref{tab:solardefault} and may be modified by the user.
            
        To convert from \((l,\,b,\,d)\) coordinates to \((x,\,y,\,z)\), {\sc SynthPop} uses the following transformation: 
        \begin{equation}
            \begin{pmatrix}
            x\\y\\z                
            \end{pmatrix} = 
            \matr{R_{y}}(-\arctan(z_{\odot}/x_{\odot}))
            \left(
            \begin{pmatrix}
            d \cos{l} \cos{b} \\ d \sin{l} \cos{b} \\ d \sin{b} 
            \end{pmatrix} 
            - \begin{pmatrix}
            d_{GC}\\ 0 \\0 
            \end{pmatrix}
            \right),
        \end{equation}
        where \(\matr{R_y}(\theta)\) is the rotation matrix around the y-axis, used to correct for the slight tilt of \(\sim0.1^{\circ}\) between the coordinate systems due the Sun's location slightly above the Galactic plane (see figure \ref{fig:cartesian_coordinates}). 

        The relevant rotation matrices are:
        \begin{equation}
            \matr{R_{x}}(\theta) = \begin{bmatrix} 1 & 0 & 0\\ 0 &\cos\theta & -\sin\theta \\ 0 & \sin\theta & \cos\theta \end{bmatrix}
            ;\quad
            \matr{R_{y}}(\theta) = \begin{bmatrix} \cos\theta & 0 & -\sin\theta \\ 0 & 1 & 0 \\ \sin\theta & 0 &\cos\theta  \end{bmatrix}
            ;\quad
            \matr{R_{z}}(\theta) = \begin{bmatrix} \cos\theta & -\sin\theta & 0 \\ \sin\theta & \cos\theta &0 \\ 0 & 0 & 1 \end{bmatrix}.
        \end{equation}

        The cartesian Galactic velocity frame $(u,v,w)$ corresponds to the cartesian coordinate frame as:
        \begin{equation}
            \begin{pmatrix} u \\ v \\ w \end{pmatrix} = 
            \begin{pmatrix} \dot{x} \\ \dot{y} \\ \dot{z} \end{pmatrix}. 
        \end{equation}
        To convert between Galactic velocities ($u$, $v$, $w$) and Galactic proper motion plus radial velocity \((v_{r},\, \mu_l,\, \mu_{b})\), we first translate the Galactocentric velocities into Heliocentric velocities:
        \begin{equation}
            \begin{pmatrix}
                u_{hc}\\ v_{hc} \\ w_{hc}
            \end{pmatrix} =
            \matr{R_{y}}(\arctan(z_{\odot}/x_{\odot})
            \left(\begin{pmatrix}
                u\\ v \\ w
            \end{pmatrix} - \begin{pmatrix}
                u_{\odot}\\v_{\odot}\\w_{\odot}\\
            \end{pmatrix}\right).
        \label{eq:rot_matrix}
        \end{equation}
        Next, we transform from Cartesian heliocentric velocities into heliocentric proper motion and radial velocity \citep{Bovy2011}:
        \begin{equation}
            \label{eq:vel_trans}
            \begin{pmatrix}
                v_{r} \\ \kappa \,d\, \mu_{l^{*}}\\  \kappa \,d\, \mu_{b}
            \end{pmatrix}
            = \matr{R_{y}}(-b) \cdot \matr{R_{z}}(-l) 
            \begin{pmatrix}
                u_{hc}\\ v_{hc} \\ w_{hc}
            \end{pmatrix},
        \end{equation}
        where \(\kappa = 4.74047~{\rm mas~yr^{-1} * km~s^{-1}~kpc^{-1}}\) is the conversion factor between proper motion times distance and velocity.
    
    \subsection{Cylindrical coordinate system}
        Many density and kinematic profiles are symmetric under rotation around the z-axis of the Galactocentric coordinate system.
        Hence, we use cylindrical coordinates \((r,\,\phi,\,\tilde{z})\) when estimating stellar densities. 
        In our convention (see Figure \ref{fig:cartesian_coordinates}, right panel),
        \(\phi\) increases in the direction of Galactic rotation, with its zero point in the direction of the Sun; \(r\) increases outward; and  \(\tilde{z}\) increases upward from the Galactic North pole, with its zero point in the Galactic plane \citep[see, e.g.][]{Sparke2007}.
        Note that this is a left-handed coordinate system.

        \subsubsection{Correction for Warp of the Milky Way} \label{sec:warpcor}
        Due to the warp of the Galaxy, the outskirts of the Galactic plane are not aligned with the $x-y$ plane of the Cartesian coordinate system. This can optionally be accounted for in the cylindrical coordinate system, resulting in \(z\ne\tilde{z}\).
    
        The warp height (\(z_{w} = z-\tilde{z}\)), can be described by a  power-law  model \citep{Chen2019}:
        \begin{equation}
           z_{w}(r,\phi) = a_{w}\cdot \left(\frac{\max(r-R_{w}; 0\,\mathrm{kpc})}{1 \mathrm{kpc}}\right)^{b_w} \sin(\phi - \phi_{w}),
        \end{equation}
        with the onset radius \(R_{w}=7.72\,\mathrm{kpc}\), the angle for the line of nodes \(\phi_{w}=17^{\circ}\), the amplitude of the warp \(a_{w}=0.060\), and exponent of the power law \(b_{w}=1.33\). These values may be altered by a user via configuration file.
        
        This produces the following transformations between \((x,\,y,\,z)\) and \((r,\,\phi,\,\tilde{z})\):
        \begin{equation}
        \begin{split}
        r &= \sqrt{x^{2}+y^{2}}; \\
        \phi &= -\arccos(x/r); \\
        \tilde{z} &= z - z_{w}(r,\,\phi);
        \end{split}
        \quad\quad
        \begin{split}
        x &= -r\cos{\phi}; \\
        y &= -r\sin{\phi}; \\
        z &= \tilde{z} + z_{w}(r,\,\phi).
        \end{split}
        \end{equation}
        The no-warp case equates to $z_w(r,\phi) = 0$, and thus $z=\tilde{z}$.

\section{Charon Interpolator}
\label{apx:charon}
    
In a standard Lagrange interpolator, a bilinear interpolation of some quantity of interest $f$ over adjacent metallicity ([Fe/H]) and age $\tau$ tracks would follow:
\begin{equation}
   f(m_{\rm init}, \tau, [{\rm Fe/H}]) = w\srm{i,j} f\srm{i,j}(m_{\rm init}) + w\srm{i,j+1} f\srm{i,j+1}(m_{\rm init}) + w\srm{i+1,j} f\srm{i+1,j}(m_{\rm init}) + w\srm{i+1,j+1} f\srm{i,j+1}(m_{\rm init}),
\end{equation}
i.e., a weighted mean. Here, $m_{\rm init}$ is the initial stellar mass, \(w_{i,j}\) are the normalized weights for each isochrone, and \(f_{i,j}\) is the cubic interpolation of $f$ in initial mass, i.e., the bin). The indices $i$ and $j$ indicate the metallicity - age grid points, where only the two nearest the target metallicity and age are summed over. 

\begin{figure}
        \centering
        \includegraphics[width=\textwidth]{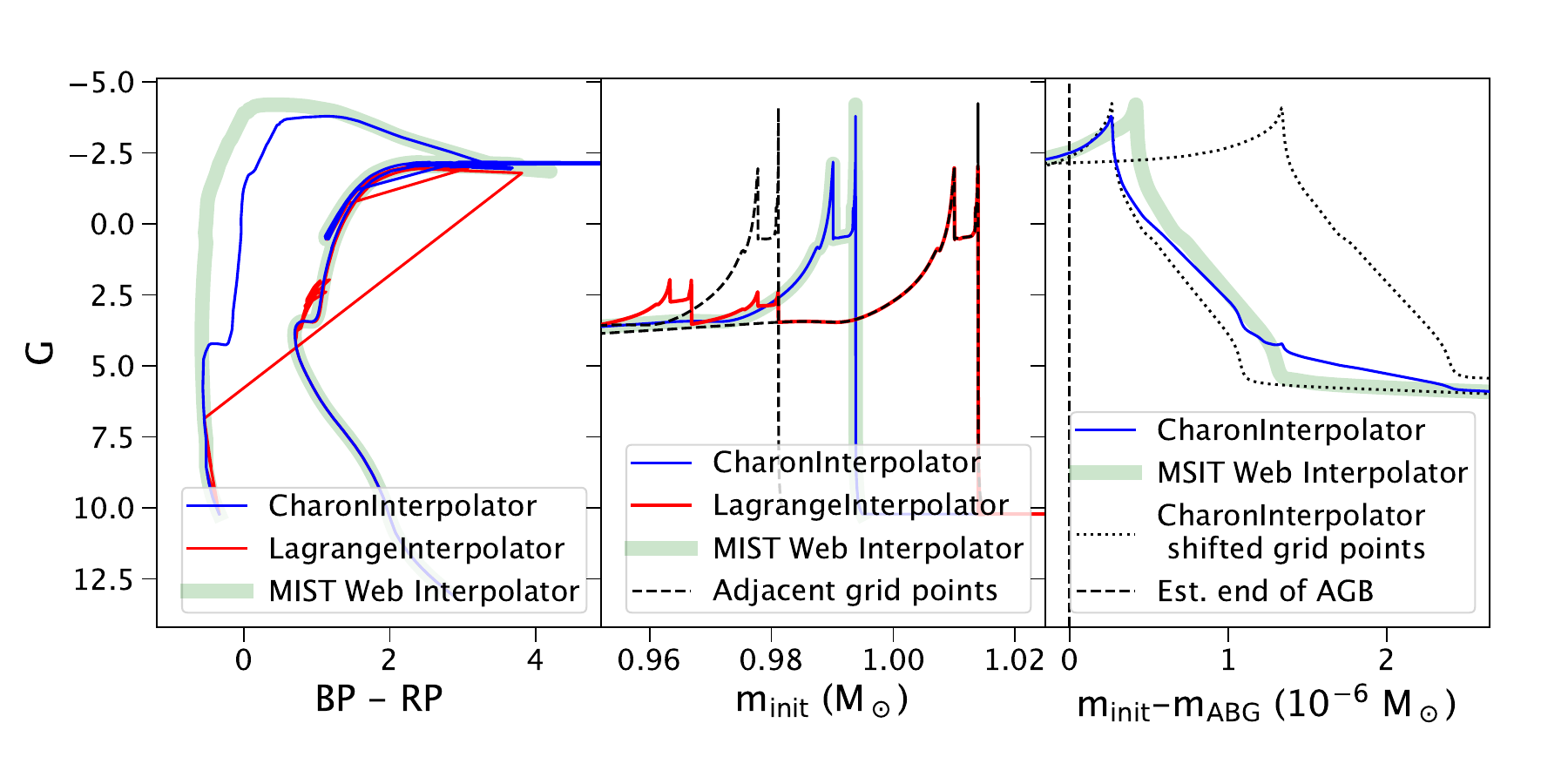}
        \caption{Color-magnitude diagram (left), initial mass-magnitude relation (center), and initial mass-magnitude diagram zoomed-in on the post-AGB phase (right) for an isochrone at a grid point metallicity and halfway between grid points in age.
        The {\it green} line shows an isochrone using the MIST Web interpolator, and the {\it red} line is the resulting isochrone from our standard interpolation scheme. For stars beyond the main sequence, this can produce spurious magnitudes and parameters. The {\it blue} line shows the isochrone resulting from the \class{CharonInterpolator} method described in Section \ref{apx:charon}. This minimizes spurious output, with a smaller effect only in the post-AGB phase. 
        In the center panel, the {\it dashed, black} lines show the isochrones for the assigned metallicity and adjacent age grid points. In the right panel, the {\it dashed, black} line marks the end of the AGB for the \class{CharonInterpolator} isochrone, and the {\it dotted, black} lines mark the isochrones for the adjacent grid points following the mass-modification scheme used by the \class{CharonInterpolator}.}
        \label{fig:comp_interp}
    \end{figure}

However, the phases of stellar evolution can be very different in consecutive mass and metallicity grid points, which can lead to inaccurate results where stars appear to bridge the gap between phases. 
The central panel of Figure \ref{fig:comp_interp} shows values for Gaia $G$ magnitude as a function of the initial mass given the standard interpolation, and the left panel shows the resultant color-magnitude diagram. 
When compared to the output of the MIST web interpolator\footnote{\url{https://waps.cfa.harvard.edu/MIST/interp_tracks.html}}, the Lagrange interpolator produces unphysical results during the transition from the red giant phase to the white dwarf phase.

Importantly, we see that the interpolation process must be tweaked for post-main sequence evolution. Problems arise when there are differences in phase between adjacent grid points. The standard way to resolve these problems in isochrone interpolation is to interpolate using the more granular Equivalent Evolutionary Phase \citep[EEP;][]{mist1} as the independent variable, but to do so requires iteratively solving for the EEP of the desired point, which we avoid for the sake of computational efficiency. Instead, we interpolate among entries with different initial masses but that are in the same evolutionary phase, and ideally the same EEP. This is implemented in the \class{CharonInterpolator} class,\footnote{In Greek mythology Charon is the ferryman that brings the dead across the river Styx. Here it handles the transition from the main sequence to their final phase.} which uses a modified initial masses $\tilde{m}_{{\rm init},i,j}$ to evaluate improved interpolated quantities of interest $\tilde{f}_{i,j}$ on the adjacent metallicity and age grid points as
\begin{equation}
\tilde{f}_{i,j}(m_{\rm init}, [{\rm Fe}/{\rm H}], \tau) = f_{i,j}(\tilde{m}_{{\rm init},i,j}(m_{{\rm init},i,j}, [{\rm Fe}/{\rm H}], \tau), [{\rm Fe}/{\rm H}], \tau),
\end{equation}
when interpolating stars in their evolved phases.
For \(\tilde{m}_{{\rm init},i,j}(m_{\rm init}, [Fe/H], \tau)\), we use a piece-wise linear function. One linear part maps the initial masses corresponding to the tip of the Red Giant Branch (\(m\srm{RGB}\)) and Asymptotic Giant Branch (\(m\srm{AGB}\)) for a given age and metallicity to the tips of the Red Giant Branch (\(m_{{\rm RGB},i, j}\)) and end of  asymptotic 
giant branch (\(m_{{\rm AGB},i,j}\)) for the adjacent age-metallicity grid point.
The same process is performed between the Terminal Age Main Sequence (TAMS) and the tip of RGB, and between the end of the AGB phase and last point on the isochrone \((m\srm{end})\). 
This is calculated via:
%

\begin{equation}
\tilde{m}_{{\rm init},i,j} = \left\{
\begin{array}{llll}
 & \frac{m\srm{init}-m\srm{TAMS}}{m\srm{RGB}-m\srm{TAMS}}(m_{{\rm RGB},i,j}-m_{{\rm TAMS},i,j}) + m_{{\rm TAMS},i,j}, & \quad & m\srm{init} < m\srm{RGB}; \\
 & \frac{m\srm{init} - m\srm{RGB}}{m\srm{AGB}-m\srm{RGB}} (m_{{\rm AGB},i,j}-m_{{\rm RGB},i, j}) + m_{{\rm RGB},i, j}, 
\quad & \quad m\srm{RGB} < & m\srm{init} < m\srm{AGB}; \\
 & \frac{m\srm{init} - m\srm{AGB}}{m\srm{end}-m\srm{AGB}} (m_{{\rm end},i,j}-m_{{\rm AGB},i, j}) + m_{{\rm end},i, j}, 
\quad & \quad m\srm{AGB} < & m\srm{init}.
\end{array} \right.
\end{equation}

The masses \(m_{{\rm TAMS},i, j}\),  \(m_{{\rm RGB},i, j}\) and \(m_{{\rm AGB},i, j}\) are extracted from the input isochrone. Within the MIST isochrone system, these are indicated by the EEP number equal to 454 (TAMS),  605 (start of RGB) and 1409 (start of AGB), respectively.  \(m_{{\rm end},i, j}\) is the last data point and corresponds to the EEP number 1710. These are interpolated linearly in [Fe/H] and $\log(\tau)$ to estimate \(m\srm{AGB}\), \(m\srm{RGB}\), \(m\srm{TAMS}\) and \(m\srm{end}\). 

To ensure a continuous transition to the main sequence, where mass doesn't need to be modified, we use:
\begin{equation}
     \tilde{m}_{i,j}' = (1-w) \cdot m\srm{init} + w \cdot \tilde{m}_{i,j},
\end{equation}
where $w$ is a weight defined as:
\begin{equation}
    w = \frac{m\srm{init} - 0.95\,m\srm{TAMS}}{0.05\,m\srm{TAMS}},
\end{equation}
and truncated to the interval [0, 1].

Finally, for each adjacent grid point, the stellar properties are estimated by interpolating the isochrone data using a cubic polynomial in the initial mass.  
This interpolation is shown as blue line in Figure \ref{fig:comp_interp}.

The modified mass interpolation method matches the results of the web interpolator for most of the evolutionary path.
However, a clear offset can be seen in the CMD during the transition from the post-AGB phase to the white dwarf cooling track. 
This is caused by a non-perfect alignment of the evolutionary tracks, as shown in the right panel. 

Although the phases are aligned in mass at the end of the AGB phase, the isochrones diverge quickly.
The result could be improved by including secondary EEPs as fixed points for the mapping.  
This would increase the complexity and the computation time, since the complete function \(m\srm{init}(EEP, [{\rm Fe}/{\rm H}], \tau)\), which is used to generate the MIST isochrones, 
must be inverted for every metallicity and age.
The likelihood for a star to be within this short transition phase is extreme low.
Therefore, we only use the primary phase transition points which keep {\sc SynthPop} more efficient and flexible in its ability to include additional isochrone systems in the future. 

\end{document}